\documentclass[a4paper,fleqn,usenatbib]{mnras}
\usepackage[T1]{fontenc}
\usepackage{ae,aecompl}
\usepackage{graphicx}
\usepackage{amsmath}
\usepackage{amssymb}

\title[Study of Planetary Nebulae PB 1 and PC 19]{Morphology and Ionization Characteristics of Planetary Nebulae PB 1 and PC 19}

\author[Rahul Bandyopadhyay et al.]{
Rahul Bandyopadhyay,\thanks{E-mail: rahul@bose.res.in}
Ramkrishna Das,
Soumen Mondal and
Samrat Ghosh
\\
S. N. Bose National Centre for Basic Sciences, Block JD, Sector III, Salt Lake, Kolkata 700106, India\\
}

\date{Accepted XXX. Received YYY; in original form ZZZ}

\pubyear{0000}

\begin{document}
%\label{firstpage}
\maketitle

\begin{abstract}
We present results of our study of two planetary nebulae (PNe), PB1 and PC 19. We use the optical spectra of these two PNe observed at 2 m Himalayan Chandra Telescope and also archival and literature data for the study. We use the morphokinematic code SHAPE to construct 3D morphologies of the PNe and the photoionization code CLOUDY to model the observed spectra. The 3D model of PB 1 consists of an elongated shell surrounded by a bipolar halo and that of PC 19 consists of an open lobed bipolar structure and a spiral filamentary pair. We analyze the ionization structure of the PNe by deriving several plasma parameters and by photoionization modelling. We estimate the elemental abundances of the the elements, He, C, N, O, Ne, S, Ar, and Cl, from our analysis. We find He, C and N abundances to be significantly higher in case of PB 1. We estimate different physical parameters of the central stars, namely effective temperature, luminosity and gravity, and of the nebula, namely hydrogen density profiles, radii, etc., from photoionization modelling. We estimate distances to the PNe as $\sim$4.3 kpc for PB 1 and as $\sim$5.6 kpc for PC 19 by fitting the photoionization models to absolute observed fluxes. Progenitor masses are estimated from theoretical evolutionary trajectories and are found to be $\sim$1.67 and $\sim$2.38 $M_{\sun}$ for PB 1 and PC 19, respectively.  
\end{abstract}


\section{Introduction}

Planetary nebulae (PNe) form towards the end of the life cycle of low- and intermediate-mass ($\sim$1-8 $M_{\sun}$) stars. The basic formation mechanism of PNe could be explained by the Interacting Stellar Wind (ISW) model \citep{1978ApJ...219L.125K}. According to this model, a circumstellar envelope, moving slowly ($v\sim10$ km s$^{-1}$) outward, is formed during the asymptotic giant branch (AGB) phase. Then it is swept by another fast ($v\sim1000$ km s$^{-1}$) wind driven by the central star during the post-AGB phase. This hydrodynamical interaction forms a compressed spherical shell around the star. The central star gradually attains high temperatures ($\sim$20000-200000 K) during its evolution towards white dwarf (WD) and photoionizes the shell. High-resolution observations reveal that majority of PNe exhibit non-spherical circumstellar shells of very complex morphologies, e.g., elliptical, bipolar, multipolar etc., with the presence of small-scale structures like knots, filaments and ansae (e.g., \citealt{2011AJ....141..134S}; \citealt{2016ApJ...830...33S}). The observed PN morphology is a result of combination of different physical processes having occurred at different points of time, few during the evolution of the central star and others within the nebular gas that evolves in dynamical time-scale. To understand the morphologies, one need to reconstruct three-dimensional (3D) models using the available observational constraints, which include the spatially resolved images and spectra correlating the position and velocity (e.g., \citealt{2006RMxAA..42...99S}; \citealt{2014ApJ...787...25H}; \citealt{2015A&A...582A..60C}).

Spectra of PNe are enriched with strong recombination lines (RLs) of H and He in emission over all wavelength regimes. Strong collisionally excited lines (CELs) of C, N, O, Ne, Ar, S, and Cl are frequently observed along with their much RLs also. Infrared spectra of PNe show the presence of molecules, such as, crystalline silicate, graphite, fullerene, PAH molecules, water masers, HCN etc. (e.g., \citealt{2017IAUS..323..141Z} and references therein). Spectral lines have been used to calculate abundances of the elements present in PNe using the direct method based on the determination of the electron temperature and the use of ionization correction factors (ICFs) and from photoionization modelling (e.g., \citealt{2017IAUS..323...43M} and references therein). In low- and intermediate-mass stars, the abundances of the elements O, Ne, S, Ar, and Cl mostly remain unchanged throughout the evolutionary process. Hence, the abundances of these elements represent that of the birth environment of the progenitor. However, He, C, and N are enriched during the evolutionary process. Thus, their abundances in PNe provide clue on tracing back the nucleosynthesis history of the progenitor (e.g., \citealt{2018MNRAS.473..241H} and references therein). The composition of PNe also predicts enrichment of the interstellar medium, as the nebulae gradually diffuse into it.

\begin{table*}
\centering
\small
\caption{Log of HCT observations. \label{tab:logobs}}
\begin{tabular}{l c c c c}
\hline
Object & Instrument/Grism & Exposures (s) & Slit orientation & Date of Observation\\
\hline
PB 1 & HFOSC/Gr. 7 & 1800 & E-W & Mar. 3, 2017\\
& HFOSC/Gr. 8 & 1800 & E-W & Mar. 3, 2017\\
& HFOSC/Gr. 7 & 1800 & E-W & Oct. 7, 2018\\
& HFOSC/Gr. 8 & 1800 & E-W & Oct. 7, 2018\\
PC 19 & HFOSC/Gr. 7 & 600$\times$2 and 300 & N-S & Jul. 8, 2018\\
& HFOSC/Gr. 8 & 300$\times$3 & N-S & Jul. 8, 2018\\
\hline
\end{tabular}
\end{table*}

Earlier several attempts have been taken to estimate the total number of observable galactic PNe and different catalogues have been prepared, for example `Strasbourg-ESO Catalogue of Galactic Planetary Nebulae' \citep{1992secg.book.....A} that included a list of about 1200 PNe, and `Version 2000 of the Catalogue of Galactic Planetary Nebulae' \citep{2001A&A...378..843K} that listed $\sim$1500 PNe. In the last decade, more sensitive surveys such as MASH (Macquarie/AAO/Strasbourg H$\alpha$ Planetary Galactic Catalogue; \citealt{2006MNRAS.373...79P}), MASH-II \citep{2008MNRAS.384..525M}, and IPHAS (Isaac Newton Telescope Photometric H$\alpha$ Survey of the Northern Galactic Plane; \citealt{2005MNRAS.362..753D}) were conducted, along with several follow-up diagnostics. From these surveys many new PNe have been detected and consequently the total number of galactic PNe (including suspected ones) has been increased to $\sim$3500. However, out of these, a few have been individually studied in detail so far (e.g. NGC 7027, NGC 6720, NGC 6853, etc.). The information acquired from the well-studied PNe in combination with the data and information collected from the observation and study of the less known PNe could play an important role in understanding the properties that may still be hidden among the PNe.

In this paper, we present the study of the characteristics of two such PNe, PB 1 and PC 19, which have not been studied individually in detail. We use optical spectra using 2 m Himalayan Chandra Telescope (HCT) and relevant archival data. We have modelled the spectra using SHAPE and CLOUDY for morphological and photoionization analyses, respectively. We have discussed about the details of data set used in this work in Section 2. The modelling procedure and results are discussed in Sections 3 and 4, respectively. We summarize and discuss the results in Section 5.

\section{Data Set}

\subsection{2 m HCT Observations and Data Reduction}

PB 1 and PC 19 were observed using the 2 m HCT located at Hanle, India, operated by Indian Institute of Astrophysics, India. Optical spectra were obtained using the Hanle Faint Object Spectrograph Camera (HFOSC) through the grisms, Gr. 7 ($\sim$3700-7000 {\AA}, $R\sim1400$) and Gr. 8 ($\sim$5500-9000 {\AA}, $R\sim2200$), hereafter referred as blue spectra and red spectra, respectively (Table \ref{tab:logobs}). A slit of dimension $1^{\prime\prime}.92 \times 11^{\prime}$ was placed approximately through the centre of the objects during observations. FeAr and FeNe lamp spectra for the wavelength calibrations to blue and red spectra, respectively, were taken immediately after each object exposure. Feige 66 ($V=10.5$, Spectral type: sdO) and Feige 110 ($V=11.8$, Spectral type: DOp) were observed as optical standards, during the observations of PB 1 and PC 19, respectively, for the flux calibration.

Spectra are reduced following the standard procedure using different tasks under the IRAF package. All spectra are, first, bias subtracted and cosmic-ray corrected, and then the 1D spectra summed across the aperture are extracted using APALL. Wavelength calibration in each spectrum is made using the corresponding reference lamp spectrum. Flux calibration is done using the sensitivity function determined from flux data of the standard star after making airmass correction. The spectra are corrected for interstellar extinction using

\begin{equation}
I(\lambda)=F(\lambda)10^{[c(\mathrm{H}\beta)f(\lambda)]}
\end{equation}

where, $I(\lambda)$ and $F(\lambda)$ are the intrinsic and observed fluxes, respectively, $c(\mathrm{H}\beta)$ is the logarithmic extinction at H$\beta$ and $f(\lambda)$ is the extinction curve derived by Cardelli, Clayton $\&$ Mathis \cite{1989ApJ...345..245C}. The value of $c(\mathrm{H}\beta)$ is obtained in order to match the theoretical Balmer line ratios in the dereddened spectrum. For the calculation using Equation 1, we adopt the theoretical intrinsic value of $I(\mathrm{H}\alpha)/I(\mathrm{H}\beta)=2.847$, corresponding to the electron temperature, $T_\mathrm{e}=10^4$ K and electron density, $N_\mathrm{e}=10^4$ cm$^{-3}$ \citep{2006agna.book.....O}.    

\begin{figure*}
\centering
\scalebox{0.7}[0.7]{\includegraphics{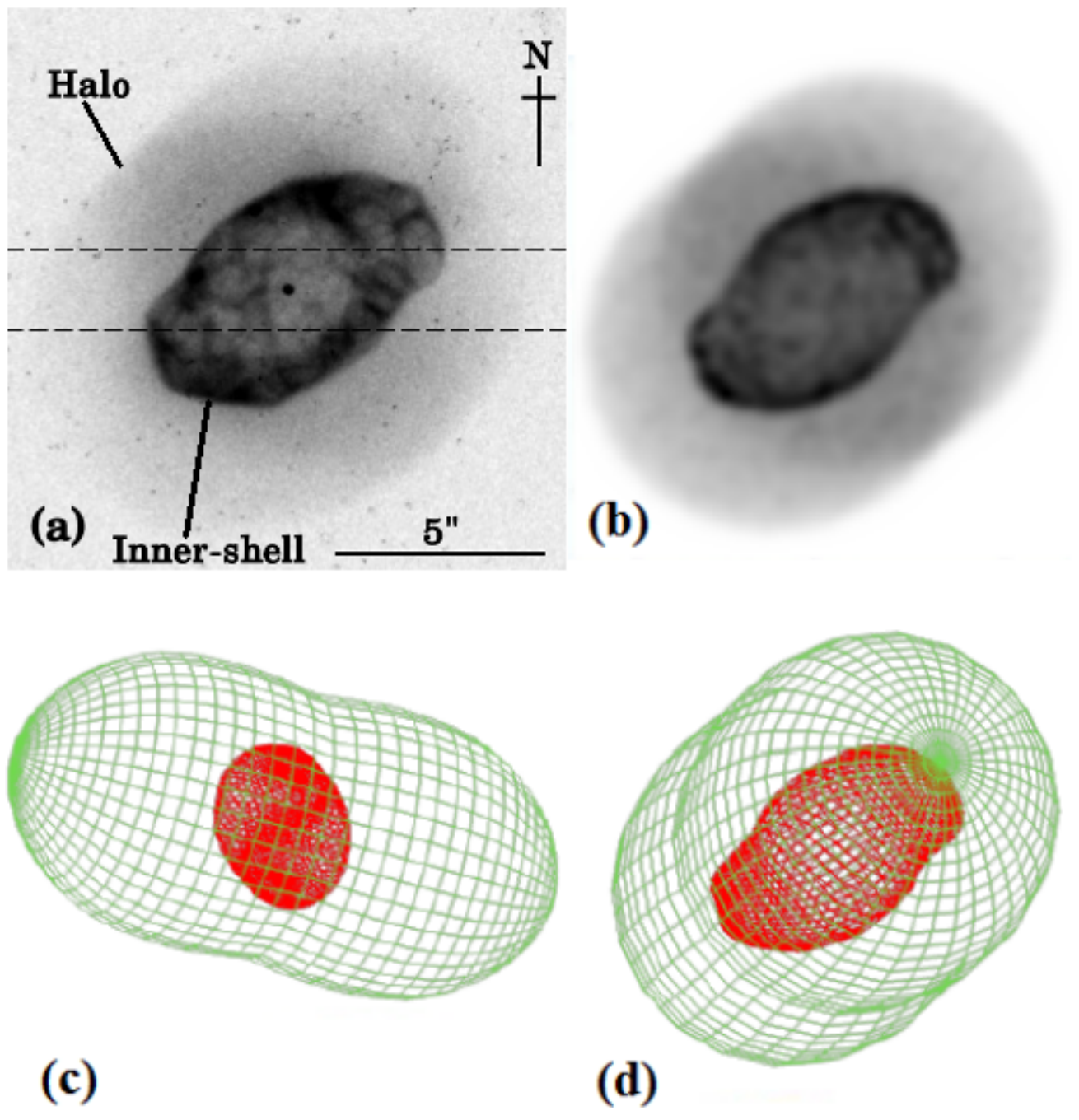}}
\caption{(a) Archival H$\alpha$ image of PB 1, used for the 3D reconstruction (see Sec. \ref{sec:hstimg} for references). The position and width of the slit used for HCT spectroscopic observation is marked with dotted line on the image. (b) The rendered grey-scale 2D model image for comparison with the observed image. (c) The side view and (d) the sky view from the Earth of the 3D model of PB 1 constructed using SHAPE.}
\label{fig:3Dmodelpb1}
\end{figure*}

\begin{figure*}
\centering
\scalebox{0.4}[0.4]{\includegraphics{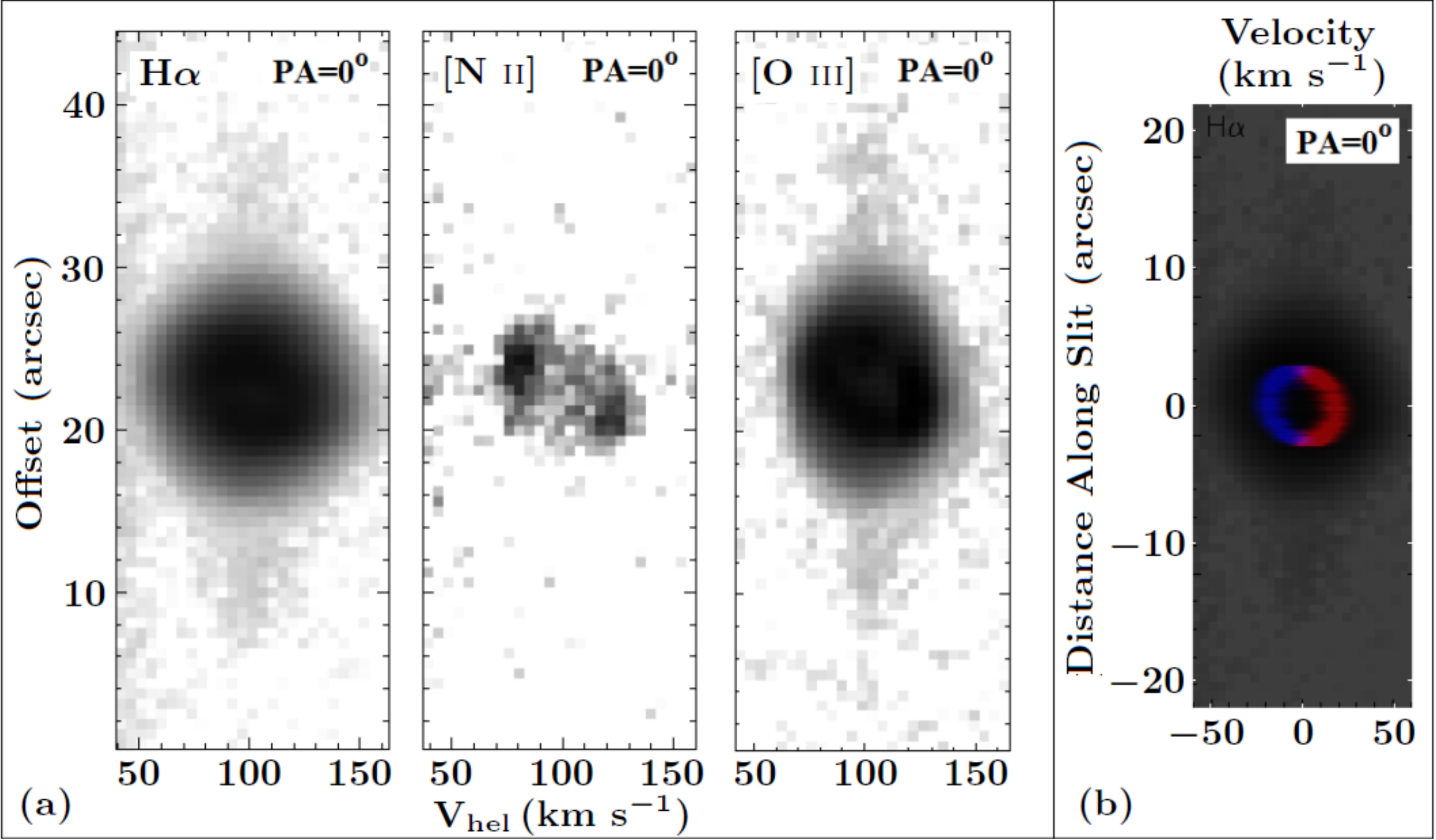}}
 \caption{(a) PV diagrams of PB 1 in H$\alpha$, [N~{\sc ii}] and [O~{\sc iii}] from \citet{SPMCatalog} used to aid the 3D modelling. All the diagrams correspond to a N-S slit-position and a slit width of $2^{\prime\prime}$. (b) Our model PV diagram (in red and blue) is superimposed on the observed PV diagram in H$\alpha$. The red and blue halves of the model PV correspond to the red and blueshifted regions, respectively.}
\label{fig:pb1pv}
\end{figure*}

\subsection{Archival Data}

\subsubsection{Narrow-band Images}  \label{sec:hstimg}

We have used pipeline-calibrated high-resolution \textit{Hubble Space Telescope (HST)} images of the PNe from the Hubble Legacy Archive (HLA) \footnote{\url{https://hla.stsci.edu/}}. The images were obtained using the Wide-Field Planetary Camera 2 (WFPC2) instrument in narrow bands at a spatial resolution of $0^{\prime\prime}.045$ pixel$^{-1}$. The image of PB 1 was taken through H$\alpha$ filter (F656N; $\lambda_c=6564\AA, \Delta\lambda=22\AA$) with an exposure of 1120 seconds on Sep. 27, 1999 (PI: Sahai; Proposal ID: 8345). The image of PC 19 was obtained with 600 seconds of exposure through [N~{\sc ii}] (F658N; $\lambda_c=6591\AA, \Delta\lambda=29\AA$) filter on Jun. 29, 1999 (PI: Borkowski; Proposal ID: 6347).\\\\

The narrow-band images of PC 19 have been taken from IAC Morphological Catalog of Northern Galactic Planetary Nebulae \citep{1996iacm.book.....M}. The observations were obtained using the 2.5 m Nordic Optical Telescope (NOT) through H$\alpha$ ($\lambda_c=6563\AA, \Delta\lambda=10\AA$), [N~{\sc ii}] ($\lambda_c=6584\AA, \Delta\lambda=10\AA$), and [O~{\sc iii}] ($\lambda_c=5007\AA, \Delta\lambda=30\AA$) filters at a spatial resolution of $0^{\prime\prime}.139$ pixel$^{-1}$.

\subsubsection{Long-slit Echelle Spectra}

Bi-dimensional, spatially resolved long-slit Echelle spectra in form of position-velocity (PV) diagrams are taken from San Pedro M\'artir Kinematic Catalogue of Galactic Planetary Nebulae \citep{SPMCatalog}. The observations for building this catalogue were carried out at San Pedro M\'artir National Observatory, Mexico, using the Manchester Echelle Spectrometer (MES). The spectra for both PB 1 and PC 19 were obtained using 2$^{\prime\prime}$ slit with resolution of 11.5 km s$^{-1}$. We have also used the long-slit Echelle spectra for PC 19 observed by \cite{1999AJ....117..967G}, on Jul. 22 and 23, 1997, using the IACUB spectrometer at the NOT, through 0$^{\prime\prime}$.64 slit at a resolution of $\simeq9-10$ km s$^{-1}$.

\section{Modeling Procedure}

\subsection{3D Modeling} \label{sec:3Dmodelling}

3D reconstruction of the nebular structure are performed using the morphokinematic modelling code SHAPE \citep{2011ITVCG..17..454S}. In SHAPE, 3D mesh model can be constructed with one or more geometrical components using the basic geometrical structures, e.g., \textit{sphere}, \textit{torus}, \textit{cone} etc. The components can be attributed with certain density profile and velocity field using \textit{density} and \textit{velocity} modifiers, respectively. The code computes radiation transfer through the nebula and generates various model outputs, such as, synthetic image (2D projection of the reconstructed model), PV spectra (PV diagrams), etc., that are compared with the observations. The basic geometry can be modified in a required way using various structural modifiers, e.g., \textit{squeeze}, \textit{spiral}, \textit{twist}, etc. The \textit{squeeze} modifier is used to obtain bipolarity and waist in a basic mesh structure by compressing it towards radially inward direction along a line. The \textit{spiral} modifier is used to obtain spiral appearance in any structure, and degree of spirality can be changed by adjusting geometric parameters. By applying \textit{twist} modifier, a structure can be twisted around any specified axis. In this work, 3D models have been constructed using mainly the high-resolution HST narrow-band images. Additionally, we apply some constrain on the model using available velocity field information of the objects in form of PV diagrams.

The comparison of the observed and model image and the PV diagrams are done using eye-estimation. At first the parameters are varied in a larger range. Afterwards the parameters are narrowed down to certain range, beyond which significant deviation from the observation has been noticed.  

\subsection{Photoionization Modeling} \label{sec:pimodelling}

For a compact and self-consistent description of the ionization structure of the nebulae, we compute individual photoionization models of PB 1 and PC 19 using the photoionization code CLOUDY (version 17.00, \citealt{2017RMxAA..53..385F} and references therein). CLOUDY uses a set of input parameters that define the ionizing source and the nebula. Generally, a spherical shell geometry is considered for the nebula with an ionizing source at the centre. The code solves ionization balance and energy balance equations at each point of the nebula, and calculate radiation transfer, generating the model spectrum at the end.

In each model computed here, the incident radiation from the central star is characterized by effective temperature and luminosity. We also consider the stellar model atmosphere, `Rauch PN Nuclei' \citep{2003A&A...403..709R} (solar abundance grid, Log $Z=0.0$), in order to include the physical effects of stellar atmosphere and include gravity of the central star as a model parameter. The shell, representing the nebula is parametrized by elemental abundances, hydrogen density, filling factor, covering factor, and inner and outer radii. The initial elemental abundances are set to the values calculated by PyNeb \citep{2015A&A...573A..42L} package (see Sections \ref{sec:pldiagpb1} $\&$ \ref{sec:pldiagpc19}). However, all the abundances have been afterwards kept as free parameters for necessary adjustments of line strengths of different elements. By defining outer radii for the nebula in the model, the ionization structure is considered as matter-bound. Filling factor is included to account for clumping effects inside the nebula.  

Since the entire nebula might not have been covered by the slit during observation, there might be some flux loss, and use of observed H$\beta$ flux as the reference value for modelling would underestimate the actual absolute fluxes. Hence, to match the flux from entire nebula, we multiply the observed spectrum by the ratio of the total H$\alpha$ flux \citep{2013MNRAS.431....2F} to the observed H$\alpha$ flux. Not to mention that the scaling is only for the purpose of obtaining the appropriate H$\beta$ flux to match with the spherical Cloudy model and relative fluxes of the emission lines to H$\beta$ would still represent that observed through the slit, and not from the entire nebula. We fit the model spectra with the this corrected spectra in absolute flux units. Fitting our photoionization model to absolute fluxes allows us to obtain an estimation of distance self-consistently from the model.   

To achieve the best-fitting model, the parameters are varied until the observables are satisfactorily reproduced by the model. To properly reproduce the ionization structure of a nebula through modelling, it is important to match line ratios governing the plasma properties (electron temperatures and densities) as close as possible to their observed values. Before adjusting the abundances to match the line fluxes of a particular element, we reproduce the ratios between different ionization states as close as possible to the observed values. The line fluxes corresponding to each element depend not only on the abundance of that element itself, but, sometimes, it is strongly influenced by abundances of other elements also. Hence, the abundance of any particular element is adjusted, when necessary, to match the line fluxes of that or other elements. An appropriate model spectrum would simultaneously match the observed continuum, the absolute H$\beta$ flux for the entire nebula, and the line fluxes relative to H$\beta$. It may be mentioned here that we do not subtract the stellar part from the observed continuum and hence, fit the total emitted continuum from the PNe. 

To test the goodness of the model, we calculate the root mean square (r.m.s.) value of the fit using the following equation, 

\begin{equation}
r.m.s.=\sqrt{\frac{1}{n} \sum_{\mathrm{i}} {\bigg[1-\frac{M_\mathrm{i}}{O_\mathrm{i}}\bigg]}^2}
\end{equation}

Here, $M_\mathrm{i}$ and $O_\mathrm{i}$ are the modelled and observed value, respectively, of the $i$th observable and $n$ denotes the total number of observables. If the r.m.s. is less than unity, then the model is considered acceptable.

We test the quality of fitting of the individual observables by calculating the quality factor, $\kappa_\mathrm{i}$, of the fitting of the $i$th observable, following the method described in \citet{2009A&A...507.1517M}. The quality factor is defined as, 

\begin{equation}
\kappa_\mathrm{i}=\frac{\mathrm{Log}(M_\mathrm{i}/O_\mathrm{i})}{\tau_\mathrm{i}} \\ \& \\ \tau_\mathrm{i}=\mathrm{Log}(1+\sigma_\mathrm{i})
\end{equation}

Here, $\sigma_\mathrm{i}$ is the error in the $i$th observable. To obtain $\kappa_\mathrm{i}$ corresponding to each emission line, $\sigma_\mathrm{i}$ in the observed emission line flux is defined as $0.1$ for $I(\mathrm{\lambda})>0.1$ $\mathrm{H}\beta$; $0.2$ for $0.1$ $\mathrm{H}\beta>I(\mathrm{\lambda})>0.01$ $\mathrm{H}\beta$ $\&$ $0.3$ for $I(\mathrm{\lambda})<0.01$ $\mathrm{H}\beta$. The fitting is considered within the tolerance limit if the absolute value of $\kappa_\mathrm{i}$ is less than 1.   

\section{Results}

\subsection{PB 1}

PB 1 (PN G226.4-03.7), discovered by \cite{1960BOTT....2s..19P}, is a galactic anticentre PN ($R.A.=07^h02^m46.8^s$, $Dec.=-13^{\circ}42^{\prime}33^{\prime\prime}.7$). Among the previous studies of PB 1, \cite{1989A&A...222..237G} estimated a Zanstra hydrogen temperature ($T_\mathrm{Z}$(H~{\sc i})) of 37500 K and Zanstra helium temperature ($T_\mathrm{Z}$(He~{\sc ii})) of 76500 K, for the central star and placed PB 1 in high-excitation class. PB 1 was classified as elliptical and point symmetric by \cite{2011AJ....141..134S} (hereafter, SMV11). Logarithmic H$\beta$ flux (in units of erg cm$^{-2}$ s$^{-1}$) was estimated to be $-12.3\pm0.4$ by \cite{1991A&AS...90...89A} and $-12.03$ by \cite{1992A&AS...94..399C}. \cite{2013MNRAS.431....2F} found a logarithmic H$\alpha$ flux of $-11.68\pm0.07$. In the following, we discuss the results obtained from our study of PB 1.

\subsubsection{Nebular Morphology: 3D Modeling} \label{sec:3dpb1}

The H$\alpha$ image of PB 1 depicts an inner shell surrounded by a halo (Figure \ref{fig:3Dmodelpb1}(a)). The overall structure is elongated approximately along the North-West (N-W) $\&$ South-East (S-E) direction. The inner shell has a dimension of $5^{\prime\prime}.1\times5^{\prime\prime}.2$. It appears to be filamentary and clumpy in a random pattern, referred as `weaves' or `mottling' by SMV11. It depicts an ``S" shaped structure, curving Westward and Eastward near the N-W and S-E regions, respectively, giving rise to the point symmetry. The origin of point symmetry seems to be due to dynamical evolution of the inner shell in a spiral motion. This is supported by the fact that density enhancements can be noticed in the upper and lower portion of the shell. The halo is limb-brightened and shows discontinuity (SMV11). It has a dimension of $10^{\prime\prime}.1\times10^{\prime\prime}.2$. A feeble density enhancement is noticed around the minor axis of the halo, seemingly at the equatorial region. Along the direction of elongation of the halo, a region with lower density appears as trough. This feature is more prominent in N-W direction than in the S-E. It is interesting to see that the curved lobes of the inner shell is approximately directed towards the trough. This suggests that the shell might have changed direction towards the lower density region in the halo. However, only 3D morphological reconstruction can truly reveal the nebular structural details properly. For the reconstruction, we primarily refer to the HST H$\alpha$ image, along with the H$\alpha$, [O~{\sc iii}] and [N~{\sc ii}] PV diagrams (Figure \ref{fig:pb1pv}(a)) taken from \citet{SPMCatalog}. We have aimed to construct the model using possible types and combination of structural components that are commonly seen among the PNe. 

To model the inner shell, initially we have tried with different combinations and orientations of multiple shells to reproduce the observed image, because the shells seen in PNe images are often modelled with combination of multiple intersecting shells (e.g., \citealt{2014ApJ...787...25H}). However, none of them could reproduce the observed structure satisfactorily. Finally, a single shell structure is adopted to reproduce the inner shell. The \textit{size} modifier is used to transform the basic spherical shell into elliptical. Afterwards, \textit{squeeze}, \textit{twist} and \textit{spiral} modifiers are applied to achieve the point symmetric appearance of the inner shell. We use customised density profile using two \textit{density} modifiers to replicate uneven distribution of matter along the surface, as evident from the 2D image. To replicate the clumpy and filamentary nature of the inner shell in the model image, we assumed a particle distribution using the \textit{particles} option inside the shell structure.

Estimation of the inclination and expansion velocity of the inner shell is guided by the PV diagrams (Figure \ref{fig:pb1pv}(a)). The inner structure is not evidently resolved in H$\alpha$ profile, whereas, in the [O~{\sc iii}] profile, an inclined ring is evident. The [N~{\sc ii}] profile clearly shows two blobs, suggesting the inclination more clearly. Although, the available PV diagrams are not adequate to properly resolve the degeneracy in simultaneous estimation of the dimension, inclination and expansion velocity of the system, we obtain an optimized result for these parameters according to eye-estimation. From the best-fitting model, we obtain the different structural parameters of the basic elliptical shell: the major-to-minor axis of the inner shell is found to be 1.67; the inclination and position angle of the inner shell are estimated as $65^\circ$ and $313^\circ$, respectively; the ratio of outer-to-inner radius is $1.05$. However, the values of these parameters only represent those of the basic elliptical shell formed using the \textit{size} modifier. The position of the blobs in the observed [N~{\sc ii}] PV diagram suggest an expansion velocity of $\sim20$ km s$^{-1}$ along the line of sight. By adjusting the maximum expansion velocity of the inner shell to $40$ km s$^{-1}$, we obtain a good match of the observed and model PV diagram as well as the velocity along the line of sight is reproduced. The comparison between the observed and modelled PV diagram is shown in Figure \ref{fig:pb1pv}(b).

The halo is modelled with a bipolar structure with a lower degree of bipolarity. Along with the mesh, we use a volume distribution of \textit{particles} for the halo. The density distribution function is chosen to be constant throughout the structure. The density enhancement in the halo seems to be due to the superposition of the lobes of the bipolar structure. Hence, the orientation of the halo is adjusted in such a way that the density enhanced region of the observed and the modelled image superpose on one another. From the best-fitting model we find that the halo is inclined at an angle of 15$^\circ$ with the line of sight and makes a position angle of $317^\circ$, similar to the inner shell. The major-to-minor axis of the halo is found out to be 2.125. In the model, we adopt Hubble like ($v \propto r$) homologous expansion law for defining the velocity field of the halo. However, the PV diagrams cannot properly resolve the separation between the halo and the inner shell and also the bipolarity in the halo is also not evident. Due to these reasons, we do not approach to set a velocity field for the halo and hence, we do not estimate the expansion velocity of the halo in this work. We estimate the distance to PB 1 from photoionization modelling (Sec. \ref{sec:pimodelpb1}) and thereby calculate the dimensions of the components in absolute units. These results are discussed later in Sec. \ref{sec:distpb1}.

In Figure \ref{fig:3Dmodelpb1}, the observed H$\alpha$ image and the rendered synthetic image (both in grey-scale) are shown side-by-side, depicting the 2D morphology being well-reproduced by our model. The side view and sky view of the mesh model of the nebular structure are also shown. The parameters estimated from the model are summarized in Table \ref{tab:3Dparpb1}. 

\begin{table*}
\centering
\caption{Parameters of 3D morphological model of PB 1. \label{tab:3Dparpb1}}
\small
\begin{tabular}{l c c c c c}
\hline
\hline
          &          & Major/ & Position            & Inclination         & Max. Expansion  \\
Component & Geometry & Minor  & Angle               & Angle               & Velocity\\
          &          & Axis   & $(\alpha$ $^\circ)$ & $(\theta$ $^\circ)$ & (km s$^{-1}$) \\
\hline
Inner shell & Elliptical & 1.67 & 313 & 65 & 40 \\
Halo & Bipolar & 2.125 & 317 & 15 & - \\
\hline
\end{tabular}
\end{table*}

\begin{figure*}
\centering
\scalebox{0.685}[0.685]{\includegraphics{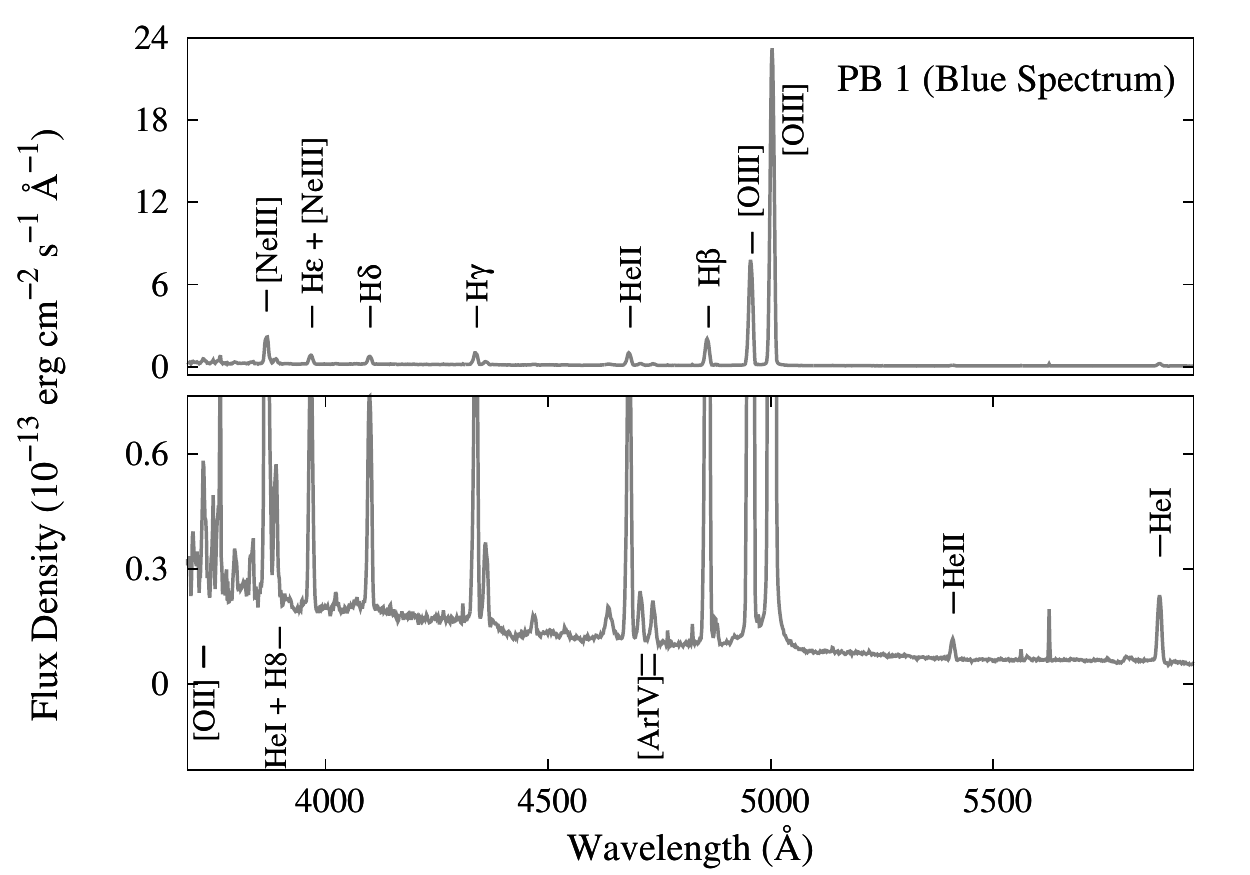}}
\scalebox{0.685}[0.685]{\includegraphics{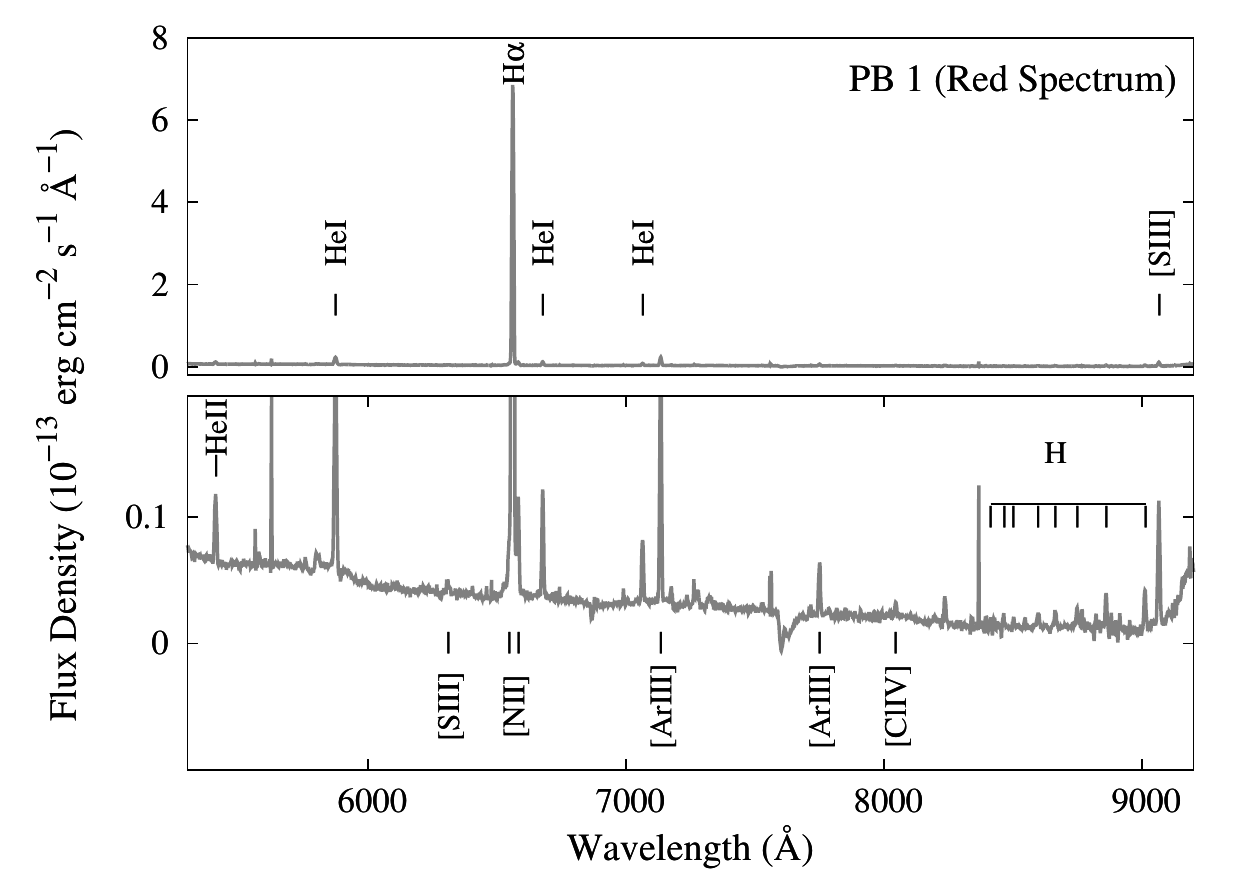}}
 \caption{The reduced dereddened blue (left) and red spectrum (right) of PB 1 obtained through Gr. 7 and Gr. 8, respectively. The spectra are zoomed vertically near the continuum to make the weaker lines visible (lower panel in both the images). Prominent emission lines are marked. Fluxes are in absolute scale. \label{fig:obsspecpb1}}
\end{figure*}

\subsubsection{Analysis of Optical Spectrum}
\subsubsection*{Emission Lines and Fluxes}
The blue and red spectrum of PB 1 are separately shown in Figure \ref{fig:obsspecpb1}. For the purpose of our analysis, we combine the two regions into one spectrum by proper scaling.
Strong RLs and CELs are present in the spectrum, suggesting PB 1 as moderately- to high-excitation PN. The commonly observed RLs of hydrogen (such as, H$\alpha$ 6563, H$\beta$ 4861, H$\gamma$ 4340, H$\delta$ 4101, and H$\epsilon$ 3970 {\AA}) and helium (such as, He~{\sc i} 4471, 5876, 6678, and 7065 {\AA}) are present in the spectrum. Presence of He~{\sc ii} 4686 {\AA} implies high central star temperature ($T_\mathrm{eff} > 50,000$).
The RL doublet C~{\sc iv} 5802, 5812 {\AA} is prominently observed. Among the CELs, [O~{\sc iii}] 5007, 4959 {\AA} and the auroral [O~{\sc iii}] 4363 {\AA}, [Ne~{\sc iii}] 3869 {\AA}, [Ar~{\sc iv}] 4711, 4740 {\AA} and [Ar~{\sc iii}] 7136, 7751 {\AA}, [S~{\sc iii}] 6312, 9069 {\AA}, [Cl~{\sc iv}] 7531, 8046 {\AA} are prominently seen. The [N~{\sc ii}] 6548 {\AA} and [N~{\sc ii}] 6583 {\AA} are partially blended with H$\alpha$ line in our spectrum. However, we cannot detect the auroral [N~{\sc ii}] 5755 {\AA} with confidence. The doublets [S~{\sc ii}] 6716, 6731 {\AA} and [Cl~{\sc iii}] 5518, 5538 {\AA} are also below confidence limit of detection.
We find the logarithmic extinction at H$\beta$, $c(\mathrm{H}\beta)=0.64$, aiming to obtain the theoretical Balmer line ratio, H$\alpha$/H$\beta$, in the dereddened spectrum. The H$\beta$ flux through slit is measured as, $F(\mathrm{H}\beta)=1.60\times10^{-13}$ erg cm$^{-2}$ s$^{-1}$.

\begin{table}
\centering
\small
\caption{Temperatures and densities for PB 1. \label{tab:specanalysispb1}}
 \begin{tabular}{lcc}
 \hline
 & This work & H10$^a$\\
 \hline
$N_\mathrm{e}$([Ar~{\sc iv}]) & ${2500}$ & -\\
$N_\mathrm{e}$([S~{\sc ii}]) & - & ${1000}$\\
$T_\mathrm{e}$([O~{\sc iii}]) & ${11409}$ & ${12170\pm694}$\\
$T_\mathrm{e}$([N~{\sc ii}]) & - & ${10300}$\\
$T_\mathrm{e}$([S~{\sc iii}]) & ${11631}$ & ${10470\pm2107}$\\
$T_\mathrm{e}$([O~{\sc ii}]) & - & ${24310\pm19090}$\\
 \hline
\multicolumn{3}{l}{$^a$Reference: \citet{2010ApJ...724..748H}}\\
 \end{tabular}
\end{table}

\begin{figure}
\centering
\scalebox{0.55}[0.55]{\includegraphics{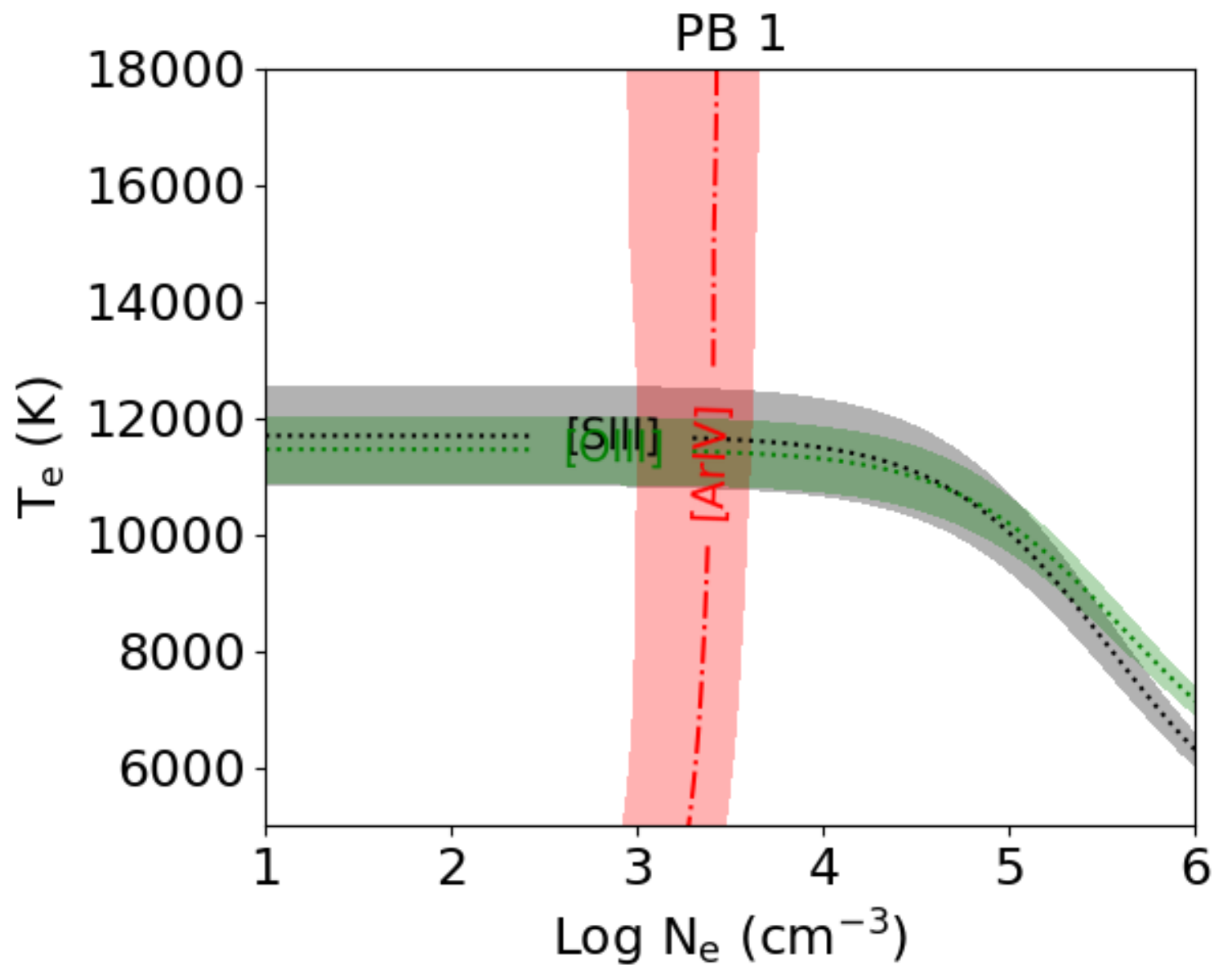}}
 \caption{The $T_\mathrm{e}$ vs Log $N_\mathrm{e}$ plot for PB 1. \label{fig:TeNepb1}}
\end{figure}

\subsubsection*{Plasma Diagnostics} \label{sec:pldiagpb1}

We analyse the physical conditions inside the nebula from the known emission line fluxes using the PyNeb package. We calculate [O~{\sc iii}], [S~{\sc iii}] electron temperatures ($T_\mathrm{e}$) and [Ar~{\sc iv}] electron density ($N_\mathrm{e}$) using the CEL ratios commonly adopted in the literature (Table \ref{tab:specanalysispb1}). As the [Ar~{\sc iv}] 4711 {\AA} is blended with He~{\sc i} 4713 {\AA}, before using [Ar~{\sc iv}] 4711 {\AA} for analysis, we have subtracted the He~{\sc i} contribution from the blended feature using the theoretical value He~{\sc i} 4713/4471$=0.146$ (considering $T_\mathrm{e}=10^4$ K and $N_\mathrm{e}=10^4$ cm$^{-3}$) \citep{1999ApJ...514..307B}. We list electron temperatures and densities estimated by \citet{2010ApJ...724..748H} for comparison with our estimated values. We also use PyNeb to obtain $T_\mathrm{e}$-$N_\mathrm{e}$ map (Figure \ref{fig:TeNepb1}) for [S~{\sc iii}], [O~{\sc iii}] and [Ar~{\sc iv}] ions to show the variation of one with the other for different ions inside the nebula. We apply the direct method to calculate the elemental abundances from the observed emission line fluxes using PyNeb. From the $T_\mathrm{e}$-$N_\mathrm{e}$ map, we approximately obtain $T_e\sim11,500$ K and $n_e\sim2500$ cm$^{-3}$, which we use during calculation of ionic abundances of the ions present in the spectrum. We further calculate the total elemental abundances for CELs of N, O, Ne and Ar, adopting the ICF expressions from \citet{2014MNRAS.440..536D}. In Table \ref{tab:modelparpb1}, we present the calculated ionic and total abundances along with the ICF values. We could not calculate S/H and Cl/H using the same method, as no [S~{\sc ii}] and [Cl~{\sc iii}] lines are observed in our spectrum. Also, we could not calculate C/H using the RL C~{\sc ii} 4267 {\AA}, as that line is not detected. 

\subsubsection{Ionization Characteristics: Photoionization Modeling} \label{sec:pimodelpb1}

\begin{table}
\centering
\small
\caption{Results of the photoionization model for PB 1. \label{tab:modelparpb1}}
\begin{tabular}{l c c}
\hline
r.m.s. && 0.13\\
H$\beta$ Flux & Log $I(\mathrm{H}\beta)$ & -11.69\\
Distance to PN & $d$ $(\mathrm{kpc})$ & 4.35\\
&&\\
\multicolumn{3}{l}{Central Star}\\
Temperature & $T_\mathrm{eff}$ $(\mathrm{K})$ & 86500\\
Luminosity & $L$ $(L_\mathrm{\sun})$ & 5646\\
Gravity & Log $g$ $(\mathrm{cm}$ $\mathrm{s^{-2}})$ & 5.0\\
&&\\
\multicolumn{3}{l}{Nebula}\\
Geometry && Closed\\
Hydrogen Density & $n_\mathrm{H}(r_\mathrm{in})$ $(\mathrm{cm^{-3}})$ & 3.65\\
Inner Radii & $r_\mathrm{in}$ $(\mathrm{pc})$ & 0.073\\
Outer Radii & $r_\mathrm{out}$ $(\mathrm{pc})$ & 0.17\\
Filling Factor & $f$ & 0.056\\
\hline
\hline
\end{tabular}
\begin{tabular}{l c c c}
\multicolumn{4}{l}{Total Elemental Abundances}\\
& Model & DM$^a$ & H10$^b$\\
\hline
He/H & 0.11 & 0.097 & 0.12\\
C/H$\times10^{4}$ & 12.11 & - & 11.0\\
N/H$\times10^{4}$ & 0.76 & 0.3 & 0.35\\
O/H$\times10^{4}$ & 3.28 & 3.63 & 3.7\\
Ne/H$\times10^{4}$ & 0.81 & 0.87 & 1.2\\
S/H$\times10^{5}$ & 0.13 & - & 0.17\\
Cl/H$\times10^{7}$ & 0.47 & - & -\\
Ar/H$\times10^{6}$ & 1.51 & 1.15 & 2.0\\
\hline
\hline
\end{tabular}
\begin{tabular}{l c c c c}
\multicolumn{5}{l}{Ionic Abundances using Direct Method}\\
& X$^{+}$/H & X$^{2+}$/H & X$^{3+}$/H & ICF\\
\hline
He	&	0.0588	&	0.0385	&	-		& 1.0\\
C	&	-		&	-		&	-	    & -\\
N	&	5.21(-7)	&	-		&	-		& 57.65\\
O	&	5.08(-6)	&	2.58(-4)	&	-		& 1.38\\
Ne	&	-		&	6.27(-5)	&	-		& 1.38\\
S	&	-		&	5.22(-7)	&	-		& -\\
Cl	&	-		&	-		&	2.75(-8)	& -\\
Ar	&	-		&	4.91(-7)	&	1.11(-6)	& 2.33\\
\hline
\hline
\end{tabular}
\begin{tabular}{l c c c c}
\multicolumn{5}{l}{Model Ionization Fractions weighted by $N_\mathrm{e}$}\\
& X$^{+}$ & X$^{2+}$ & X$^{3+}$ & ICF\\
\hline
He	&	0.6471	&	0.3524	&	-	&	1.0\\
C	&	0.0123	&	0.6353	&	0.3467	&	1.57	\\
N	&	0.0067	&	0.5808	&	0.4121	&	148.94	\\
O	&	0.0104	&	0.8551	&	0.1343	&	1.16	\\
Ne	&	0.0092	&	0.9290	&	0.0627	&	1.08	\\
S	&	0.0089	&	0.3999	&	0.5534	&	112.46	\\
Cl	&	0.0072	&	0.3741	&	0.6081	&	2.67	\\
Ar	&	0.0012	&	0.4178	&	0.5715	&	2.39	\\
\hline
\multicolumn{5}{l}{$^a$DM stands for `Direct Method'}\\
\multicolumn{5}{l}{$^b$Reference: \citet{2010ApJ...724..748H}}\\
\end{tabular}
\end{table}

\begin{table*}
\centering
\small
\caption{Comparison between observed and modelled line fluxes of PB 1. Observed and modelled flux values are given with respect to $I(\mathrm{H}\beta)=100$. \label{tab:obsvsmodpb1}}
\begin{tabular}{cccccc}
\hline
Wavelength ({\AA}) & Line Id. & Observed Flux & Modeled Flux & Mod./Obs. & $\kappa_\mathrm{i}$\\
\hline
3727	&	[O~{\sc	ii}]				&	14.72	&	12.12	&	0.82	&	-2.04	\\
3869	&	[Ne~{\sc	iii}]				&	99.11	&	100.10	&	1.01	&	0.10	\\
3889	&	He~{\sc	i}	,	H8		&	21.14	&	20.58	&	0.97	&	-0.28	\\
															
3968	&	[Ne~{\sc	iii}]	,	H$\epsilon$		&	32.30	&	46.53	&	1.44	&	3.83	\\
															
4026	&	He~{\sc	ii}	,	He~{\sc	i}	&	1.89	&	2.33	&	1.23	&	1.15	\\
															
4101	&	H$\delta$					&	27.96	&	26.39	&	0.94	&	-0.61	\\
4340	&	H$\gamma$					&	43.99	&	47.32	&	1.08	&	0.77	\\
4363	&	[O~{\sc	iii}]				&	11.62	&	13.67	&	1.18	&	1.71	\\
4471	&	He~{\sc	i}				&	2.99	&	3.51	&	1.17	&	0.88	\\
4541	&	He~{\sc	ii}				&	1.47	&	1.54	&	1.05	&	0.27	\\
4686	&	He~{\sc	ii}				&	45.80	&	39.37	&	0.86	&	-1.59	\\
4711	&	[Ar~{\sc	iv}]				&	6.09	&	5.51	&	0.90	&	-0.55	\\
4740	&	[Ar~{\sc	iv}]				&	5.83	&	4.16	&	0.71	&	-1.85	\\
4861	&	H$\beta$					&	100.00	&	100.00	&	1.00	&	-	\\
4959	&	[O~{\sc	iii}]				&	393.36	&	442.70	&	1.13	&	1.24	\\
5007	&	[O~{\sc	iii}]				&	1172.08	&	1320.80	&	1.13	&	1.25	\\
5411	&	He~{\sc	ii}				&	2.62	&	3.36	&	1.28	&	1.36	\\
5755	&	[N~{\sc	ii}]				&	-	&	0.08	&	-	&	-	\\
5802	&	C~{\sc	iv}				&	0.45	&		&	-	&	-	\\
5812	&	C~{\sc	iv}				&	0.41	&		&	-	&	-	\\
5876	&	He~{\sc	i}				&	9.03	&	9.69	&	1.07	&	0.39	\\
6037	&	He~{\sc	ii}				&	0.18	&	0.11	&	0.60	&	-1.93	\\
6234	&	He~{\sc	ii}				&	0.29	&	0.17	&	0.59	&	-2.00	\\
6312	&	[S~{\sc	iii}]				&	0.43	&	0.40	&	0.92	&	-0.30	\\
6548	&	[N~{\sc	ii}]				&	1.37	&	1.21	&	0.89	&	-0.66	\\
6563	&	H$\alpha$					&	273.07	&	275.90	&	1.01	&	0.11	\\
6583	&	[N~{\sc	ii}]				&	3.54	&	3.57	&	1.01	&	0.06	\\
6678	&	He~{\sc	i}				&	3.19	&	2.55	&	0.80	&	-1.23	\\
6891	&	He~{\sc	ii}				&	0.17	&	0.41	&	2.44	&	3.40	\\
7065	&	He~{\sc	i}				&	2.05	&	3.22	&	1.58	&	2.49	\\
7136	&	[Ar~{\sc	iii}]				&	8.15	&	9.91	&	1.22	&	1.07	\\
7177	&	He~{\sc	ii}				&	0.51	&	0.51	&	1.00	&	0.00	\\
7238	&	[Ar~{\sc	iv}]				&	0.41$^a$&	0.07	&	0.18	&	-6.54	\\
7281	&	He~{\sc	i}				&	0.82	&	0.55	&	0.67	&	-1.51	\\
7325	&	[O~{\sc	ii}]				&	1.16	&	1.23	&	1.06	&	0.31	\\
7531	&	[Cl~{\sc	iv}]				&	0.15	&	0.22	&	1.48	&	1.50	\\
7751	&	[Ar~{\sc	iii}]				&	1.78	&	2.35	&	1.32	&	1.52	\\
8046	&	[Cl~{\sc	iv}]				&	0.52	&	0.50	&	0.98	&	-0.09	\\
8323	&	P25					&	-	&	1.25	&	-	&	-	\\
8334	&	P24					&	-	&	0.36	&	-	&	-	\\
8346	&	P23					&	-	&	0.27	&	-	&	-	\\
8359	&	P22					&	-	&	0.25	&	-	&	-	\\
8374	&	P21					&	-	&	0.27	&	-	&	-	\\
8392	&	P20					&	0.36	&	0.30	&	0.82	&	-0.75	\\
8413	&	P19					&	0.36	&	0.34	&	0.93	&	-0.28	\\
8438	&	P18					&	0.15	&	0.39	&	2.51	&	3.50	\\
8467	&	P17					&	0.42	&	0.45	&	1.08	&	0.28	\\
8502	&	P16					&	0.26	&	0.54	&	2.07	&	2.78	\\
8545	&	P15					&	0.27	&	0.65	&	2.36	&	3.28	\\
8598	&	P14					&	0.71	&	0.79	&	1.11	&	0.39	\\
8665	&	P13					&	0.67	&	0.97	&	1.45	&	1.42	\\
8750	&	P12					&	0.77	&	1.23	&	1.60	&	1.79	\\
8863	&	P11					&	1.07	&	1.58	&	1.47	&	2.12	\\
9015	&	P10					&	1.35	&	1.82	&	1.34	&	1.61	\\
9069	&	[S~{\sc	iii}]				&	4.61	&	4.63	&	1.01	&	0.03	\\
\hline
\multicolumn{6}{l}{$^a$Flux is uncertain}\\
\end{tabular}
\end{table*}

\begin{figure*}
\centering
\scalebox{0.9}[0.9]{\includegraphics{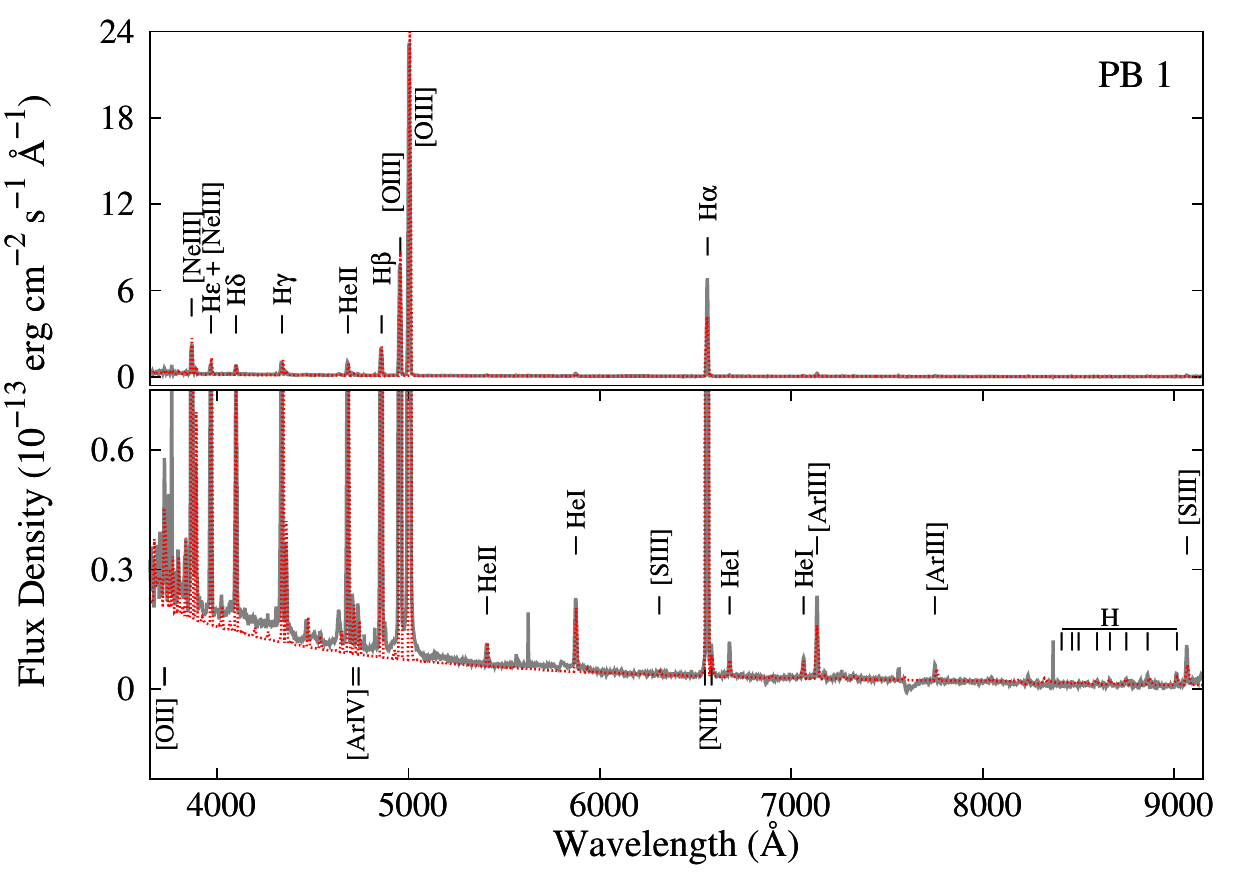}}
\caption{The observed optical spectrum (grey solid line) and the modelled spectrum (red dashed line) for PB 1 are shown. Prominent emission lines are marked. Fluxes are in absolute scale. The vertically zoomed spectrum in the lower panel shows the fit of the weaker lines and the continuum (See Section \ref{sec:pimodelpb1} for details). \label{fig:pimodelpb1}}
\end{figure*}

\begin{figure}
\centering
\scalebox{0.7}[0.7]{\includegraphics{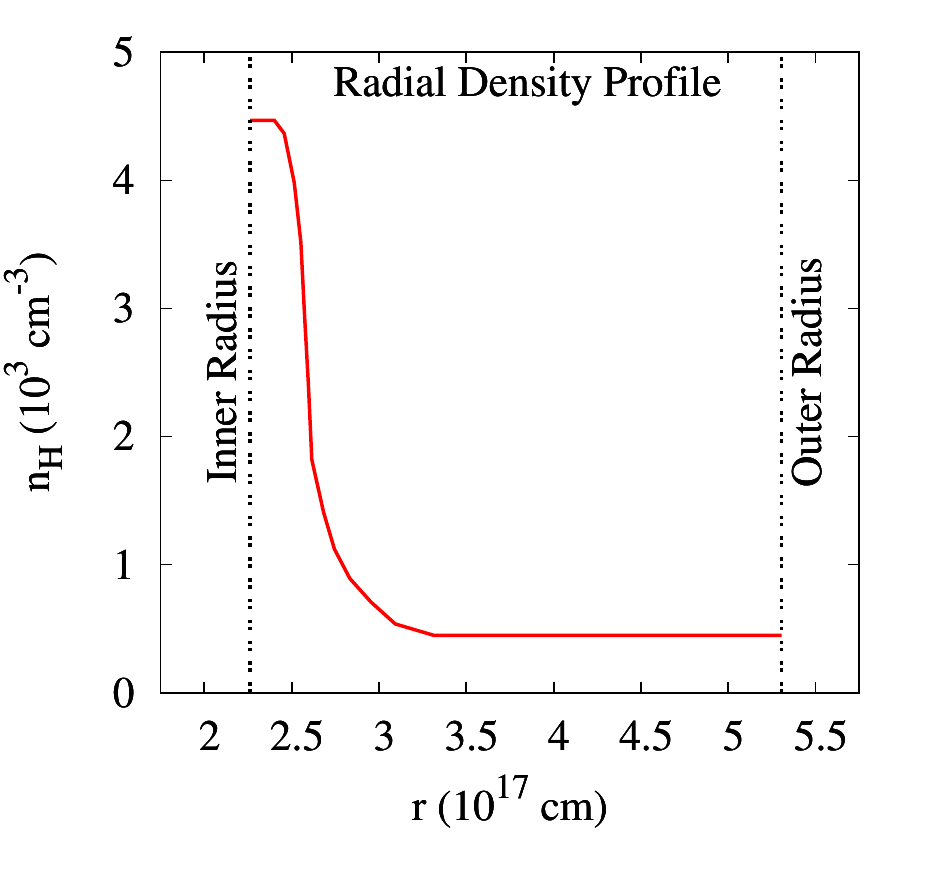}}
\caption{The hydrogen radial density profile for PB 1 used as the input of CLOUDY to compute the PI model. \label{fig:hden}}
\end{figure}

Photoionization modelling of PB 1 is performed following the basic modelling procedure described in Section \ref{sec:pimodelling}. A spherical shell geometry is considered around the central source. We assume the nebula to be matter bound by defining outer radius for the shell. As $T_\mathrm{Z}$(H~{\sc i}) (37500 K) is considerably lower than $T_\mathrm{Z}$(He~{\sc ii}) (76500 K), it seems that the nebula is not optically thick in hydrogen Lyman continuum. This supports the assumption of matter-bound case in our model. 

Figure \ref{fig:pimodelpb1} shows the fit of the observed spectrum with the model spectrum. The comparison between the observed and modelled fluxes of the emission lines are given in Table \ref{tab:obsvsmodpb1}. The quality of fit of each emission line can be verified from the calculated ratio of the modelled and the observed flux value. Though we compare all the observed line fluxes with the model line fluxes, we focus on fitting particular lines characterizing the overall ionization structure throughout the modelling and calculate the r.m.s value of the fitting using these lines. These lines include He~{\sc i} 5876 {\AA}, He~{\sc ii} 4686 {\AA}, [O~{\sc ii}] 3727, 7325 {\AA}, [O~{\sc iii}] 5007, 4363 {\AA}, [N~{\sc ii}] 6583 {\AA}, [Ne~{\sc iii}] 3869 {\AA}, [S~{\sc iii}] 9069 {\AA}, [Ar~{\sc iv}] 4711, 4740 {\AA}, [Ar~{\sc iii}] 7136 {\AA}, and [Cl~{\sc iv}] 8045 {\AA}. The r.m.s value of the model fit is calculated to be 0.13, well below unity. The value of the quality factor ($\kappa_\mathrm{i}$) corresponding to each line, as discussed in Sec. \ref{sec:pimodelling}, is included in Table \ref{tab:obsvsmodpb1}. 

We have taken care of the line ratios used for plasma diagnostics (as discussed in Sec. \ref{sec:pldiagpb1}), which include [O~{\sc iii}] 5007/4363, [S~{\sc iii}] 9069/6312 and [Ar~{\sc iv}] 4740/4711, while matching the observed spectra. In the best fit model, the first two ratios match well with the observation, though, the latter is slightly lower than the observed value. Also, we have tried to reproduce the ratios among different ionization states of same element as closely as possible in order to properly constrain the abundances. However, in our model, the ionization balance of [O~{\sc ii}]/[O~{\sc iii}] is close to the observed value and the ratio [Ar~{\sc iii}]/[Ar~{\sc iv}] remains slightly higher than the observed value. Any procedure to fit the ratio [Ar~{\sc iii}]/[Ar~{\sc iv}] with the observed value was affecting the balance of [O~{\sc ii}]/[O~{\sc iii}]. Since the abundance determination of other elements (except He) depend on the value of [O~{\sc ii}]/[O~{\sc iii}], we emphasize on the fitting of this ratio. 
 
The results of modelling are summarized in Table \ref{tab:modelparpb1}. From the best-fitting model, we obtain central star's effective temperature to be $86500$ K. The temperature is slightly above the $T_\mathrm{Z}$(He~{\sc ii}) ($76500$ K). Our model He~{\sc ii} 4686 {\AA} is slightly underestimated than the observed. We have had to keep the Log $g$ of the central star at 5.0 in order to fit the line strength of observed He~{\sc ii} 4686 {\AA} in our model. We could not test our model for lower value of Log $g$ as 5.0 is the minimum value allowed for the model atmosphere. Any other methods (such as, increasing $T_\mathrm{eff}$) to fit this line have been affecting the other observables. However, He~{\sc ii} 4686 {\AA} might also have been enhanced in the observed spectrum as the slit only partially covered the nebula towards the central region. 

We have introduced a filling factor in our model due to the evidence of clumpy nature of the inner shell in the HST image. We have also noticed that the fitting of different ionization states are also improved due to the inclusion of filling factor. However, the presence of filling factor as a free model parameter may involve degeneracy in the simultaneous estimation of luminosity, distance and nebular radii. The method applied to constrain these parameters during modelling, along with their estimated values, is discussed in Sec. \ref{sec:distpb1}. 

We have adopted a variable density profile ($n_\mathrm{H}=n_\mathrm{H}(r)$) as model input, using the \textit{density law} option in CLOUDY, to account for the structure of PB 1, which consists of high-density inner shell followed by the low-density halo. The radial density profile shown in Figure \ref{fig:hden} is used as the input hydrogen density of the nebula in our model. Similar assumption of density profile were adopted in case of other PNe also, e.g., Lin 39 \citep{2016MNRAS.462...12O} and NGC 6781 \citep{2017ApJS..231...22O}. Use of this density profile in place of a constant hydrogen density in the shell improve the overall fitting of the higher to lower ionization line ratios of same elements. In particular, we observe a sharp improvement in the fitting of observed [Ar~{\sc iv}]/[Ar~{\sc iii}] ratio. The density profile shows the hydrogen density of the inner shell to be $\sim4.5\times10^{3}$ cm$^{-3}$ and that of the halo to be $\sim4.5\times10^{2}$ cm$^{-3}$. We consider the density profile smoothly decreasing at the interface of the inner shell and halo instead of a one-step function of the radius. The assumption of smoothly varying profile seem to be more realistic than the presence of discontinuity in the density profile inside the shell. Further, it helps to avoid interruption in computation due to the abrupt change in the function.

We find that with our estimated physical parameters, we obtain a good match between the observed and modelled continuum, and also, for example, the observed absolute H$\beta$ flux, Log $I(\mathrm{H}\beta)=-11.68$, matches well with the model value of Log $I(\mathrm{H}\beta)=-11.69$.

In Table \ref{tab:modelparpb1}, we present the elemental abundances estimated from photoionization modelling and those obtained using direct method (Sec. \ref{sec:pldiagpb1}) along with the abundances obtained by \citet{2010ApJ...724..748H} (hereafter H10) for comparison. In Table \ref{tab:modelparpb1}, we also present the ionic abundances and ICF values calculated for obtaining the total elemental abundances from direct method, and the model ionic fractions of the elements along with the corresponding ICF values calculated from those fractions.

The abundances, He/H and O/H, obtained from photoionization modelling and direct method (using ICF) in this work match well with those estimated by H10. However, our model N/H is about twice compared to the H10 value and also to the value obtained using ICF in our analysis. Both of our estimated values of Ne/H from model and using ICF are slightly lower than that obtained by H10. Our Ar/H value obtained from model is slightly lower than that obtained by H10, while being slightly higher than that estimated using ICF in this work. In case of S and Cl, no information about the abundances could be obtained using direct method. Further, lines from only one ionization state of these elements are observed in the spectrum.  S/H from the model is close to that obtained by H10. We did not find any reference to compare our Cl/H value. As our photoionization model satisfactorily reproduces the observed ionization structure, we may trust and accept these abundance values.
 
We could not obtain C/H from the plasma analysis. The only C lines present in our spectrum are the RLs C~{\sc iv} 5802, 5812 {\AA}, which we could not generate using CLOUDY. Hence, we kept C/H constant at solar value ($2.69\times10^{-4}$) \citep{2010Ap&SS.328..179G} at the initial stage of modelling, which is much below the previous estimation by H10 ($11.0\times10^{-4}$). Simultaneously, we found that the model line ratio [O~{\sc iii}] 5007/4363 was remaining much less than the observed value. We could not properly reproduce the line ratio using other cooling methods. These led us to increase the C abundance in the photoionization model that acts for cooling of the plasma, and reproduces the required line ratio [O~{\sc iii}] 5007/4363 in turn. Following this method, C/H is obtained as $12.11\times10^{-4}$, which is close to that estimated by H10. We obtain C/O$=3.8$ from our result compared to C/O$=3.1$ obtained by H10.

\subsubsection{Distance, Nebular Radii and Luminosity} \label{sec:distpb1}

Previously, distance to PB 1 had been calculated as $4.3$ kpc by \citet{1998AJ....115.1989T} and $5.243\pm1.049$ kpc by \citet{2010ApJ...714.1096S}. Most recently, from Gaia DR2 parallaxes, \citet{2018AJ....156...58B} statistically determined the distance to be 3.925 kpc with a lower and an upper bound of 2.673 and 6.242 kpc, respectively.

The process of obtaining distance ($d$) using photoionization modelling, by fitting the observed spectrum in absolute fluxes, may involve degeneracy if the nebular radius ($r$), luminosity ($L$), and nebular filling factor ($f$) are also free parameters in the same model. If the model distance is appropriately estimated, the model should reproduce the observed absolute flux. Simultaneously, the angular extent of the nebula measured using the estimated distance and radii should match the angular extent of the nebula as observed in the 2D image. The readers may find a similar argument in \citet{2009ApJ...705..509O}. 

In our modelling process, first, we choose several distance values in a range of $\sim$2-7 kpc (referring the range of literature values) and calculate nebular radii ($r_\mathrm{in}$ and $r_\mathrm{out}$) for each value of distance using the 2D image and our proposed 3D morphology of the nebula. For each set of distance and corresponding radii, we run several models by varying rest of the parameters. By studying all the model generated spectra, we obtain a range of distance for the models have yielded better fit of the observed spectrum. During this process, we adopted a filling factor $\sim$0.05 and slightly varied around that value to fit the H$\beta$ flux and the ratios of higher to lower ionization lines at the final stages of modelling.

We find that the overall observed spectrum (line fluxes and the continuum) fits well for the distance approximately in the range of $\sim$3.25-3.75 kpc. Thus we adopt an average value of $d=3.5$ kpc in our final model. We estimate a luminosity of 3650 $L_{\sun}$ and a filling factor of 0.07.

However, the estimated distance may involve uncertainties as $d$, $L$ and $f$ may involve degeneracy. Theoretically, the model estimated values are degenerate with another model calculated with the values, $d=x\times3.5$ kpc (with $r_\mathrm{in}=x\times0.059$ pc and $r_\mathrm{out}=x\times0.138$ pc, along with the values $L=x^2\times3650$ $L_{\sun}$ and $f=0.07/x$ (e.g. \citealt{2009A&A...507.1517M}), where $x$ is a real number. We constrain the distance properly using the information from stellar evolutionary models and following independent method described in \cite{2009A&A...507.1517M}. By studying the `H-burning PNN evolutionary models' ($Z=0.016$) \citep{1994ApJS...92..125V}, at $T_\mathrm{eff}=86500$ K, we obtain the relation between the age ($t$) and the luminosity ($L$) of the progenitor as, 
\begin{equation}
\mathrm{Log}(t)=13.5547-2.64644\times\mathrm{Log}(L/L_{\sun})
\end{equation}
If we assume the nebular expansion velocity, $v_\mathrm{exp}=40$ km s$^{-1}$, obtained for the inner shell from the 3D model, we obtain a nebular age of 3380 yr, considering $r_\mathrm{out}=0.138$ pc as the edge of the nebula. As the nebular age is linearly proportional to distance and nebular radii, it is also degenerate and $t=x\times3380$ yr. This, along with $L/L_{\sun}=x^2\times3650$ leads to the relation,
\begin{equation}
\mathrm{Log}(t)=1.75+0.5\times\mathrm{Log}(L/L_{\sun})
\end{equation}
From equations 4 and 5, we obtain a luminosity of 5646 $L_{\sun}$, and $x=1.2437$, thereby obtain the distance of 4.35 kpc, and a filling factor of 0.056 as the final model values for these parameters. For $d=4.35$ kpc, an 1$^{\prime\prime}$.0 angular separation would correspond to $0.0211$ pc in the sky plane. Using this value, with the information about relative dimensions of the components obtained from the 3D model, we have measured the dimensions of the inner shell and the halo in absolute units (here, in pc). We find the major axis and minor axis of the inner shell (considering inner radius) to be $0.091$ and $0.054$ pc, respectively, and those of the halo to be $0.226$ and $0.106$ pc, respectively. We take the average values of the axes of the inner shell as $0.073$ pc and of the halo as $0.17$ pc and use these values as $r_\mathrm{in}$ and $r_\mathrm{out}$, respectively, in our photoionization model.

\subsection{PC 19}

PC 19 (PN G032.1+07.0), discovered by \cite{1961BOTT....3...33P}, is located towards the galactic centre ($R.A.=18^h24^m44.5^s$, $Dec.=+02^{\circ}29^{\prime}27^{\prime\prime}.9$. In the previous studies, \cite{2002ApJ...576..285S} estimated an upper limit to the central star temperature, $T_\mathrm{eff}=88920$ K (Log $T_\mathrm{eff}=4.949$) and a luminosity, $L=6577L_{\sun}$ (Log $L/L_{\sun}=3.818$) from Zanstra analysis. Logarithmic H$\beta$ and H$\alpha$ flux was obtained as $-12.3\pm0.1$ \citep{1991A&AS...90...89A} and $-11.36\pm0.04$ \citep{2013MNRAS.431....2F}, respectively. PC 19 is known to have a pair of low ionization filaments that gives a point symmetric appearance \citep{1999AJ....117..967G, 2001ApJ...547..302G}.Also, in the IAC Morphological Catalog of Northern Galactic Planetary Nebulae \citep{1996iacm.book.....M}, it has been classified as `point symmetric'. Whereas, the PN has been classified as `Spiral' by SMV11. In the following, we describe the results obtained from morphological and photoionization modelling of PC 19

\begin{figure}
\centering
\scalebox{0.45}[0.45]{\includegraphics{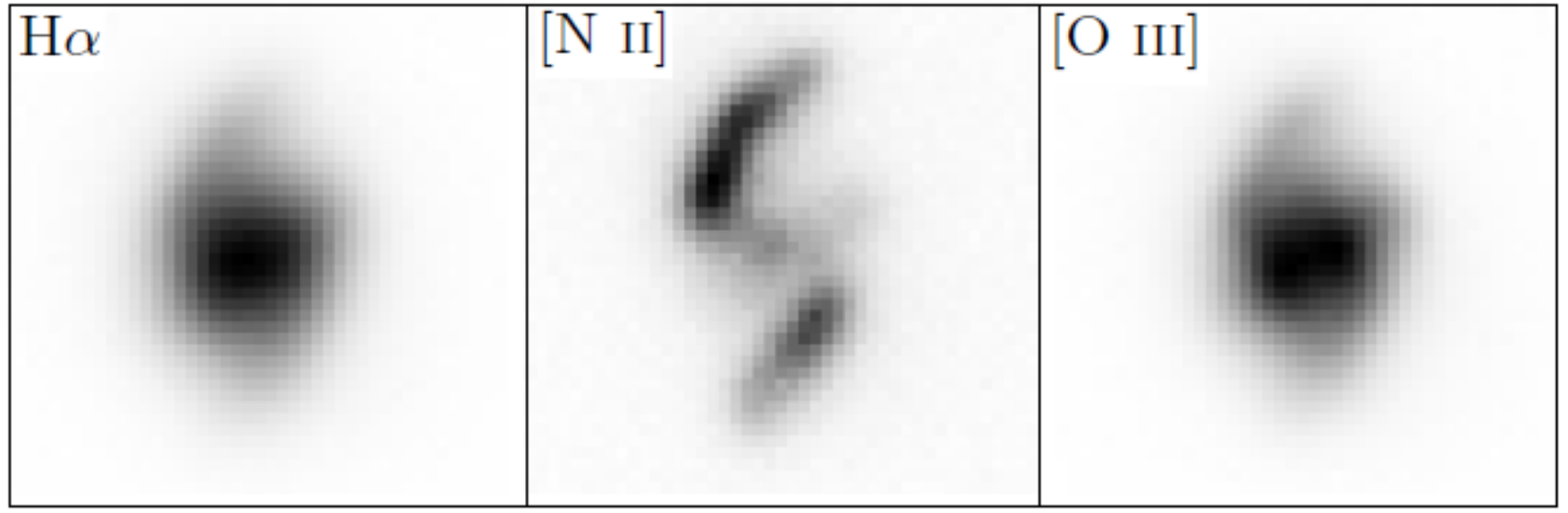}}
 \caption{Left to right: The observed images of PC 19 in H$\alpha$, [N~{\sc ii}] and [O~{\sc iii}], respectively, from IAC Catalog. (see Section \ref{sec:3dpc19})}
\label{fig:pc19iac}
\end{figure}

\begin{figure*}
\centering
\scalebox{0.7}[0.7]{\includegraphics{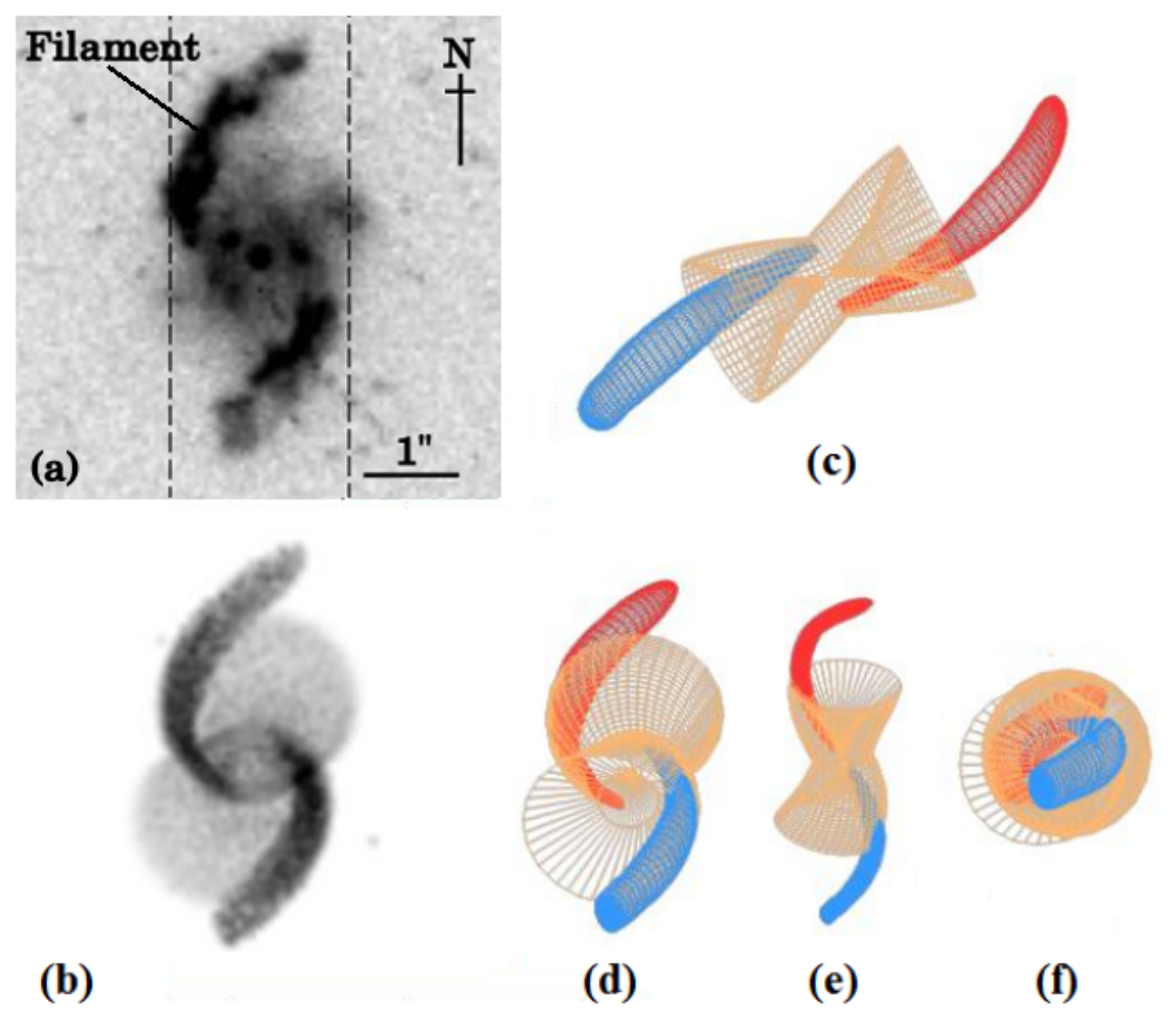}}
 \caption{(a) Archival [N~{\sc ii}] image of PC 19, used for the 3D reconstruction (see references in Sec. \ref{sec:hstimg}). The position and width of the slit used for HCT spectroscopic observation is marked with dotted line on the image. (b) The rendered grey-scale 2D model image for comparison with the observed image. The (c) side view, (d) sky view from the Earth, (e) top view and (f) front view of the 3D model of PC 19 construct using SHAPE.}
  \label{fig:3Dmodelpc19}
\end{figure*}

\begin{figure*}
\centering
\scalebox{0.4}[0.4]{\includegraphics{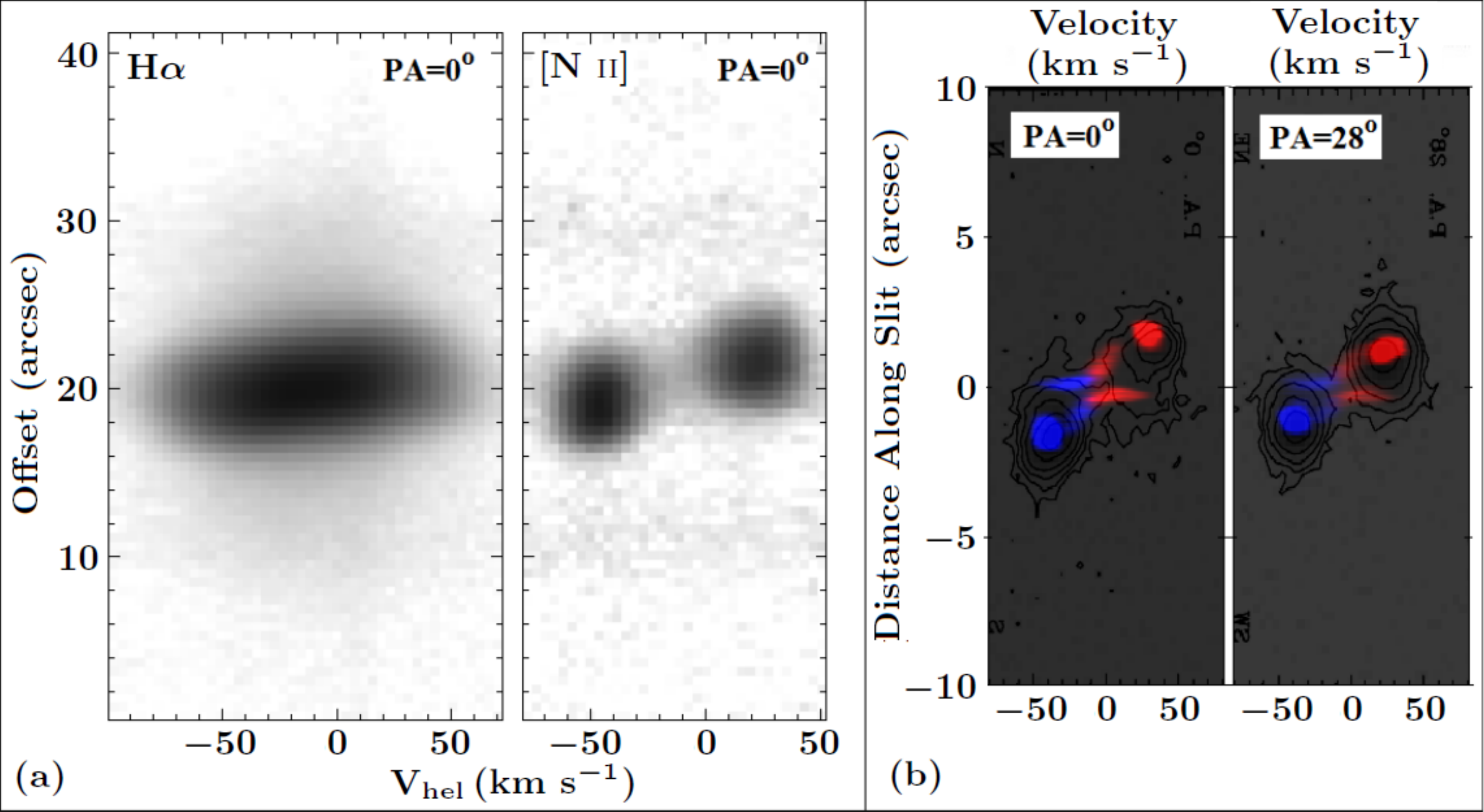}}
 \caption{
(a) PV diagrams in H$\alpha$ and [N~{\sc ii}] from \citet{SPMCatalog} used to aid the 3D modelling. All the diagrams correspond to a N-S slit-position and a slit width of $2^{\prime\prime}$. (b) The observed [N~{\sc ii}] PV diagrams corresponding to slit-positions 0$^\circ$ and 28$^\circ$ (slit width $0^{\prime\prime}.64$) taken from \citet{1999AJ....117..967G}, overplotted with our model PV diagrams obtained with identical slit orientations and slit width (in red and blue). The red and the blue halves of the model PV diagrams correspond to the red and blueshifted regions, respectively.}
  \label{fig:pc19pv}
\end{figure*}

\begin{figure}
\centering
\scalebox{0.7}[0.7]{\includegraphics{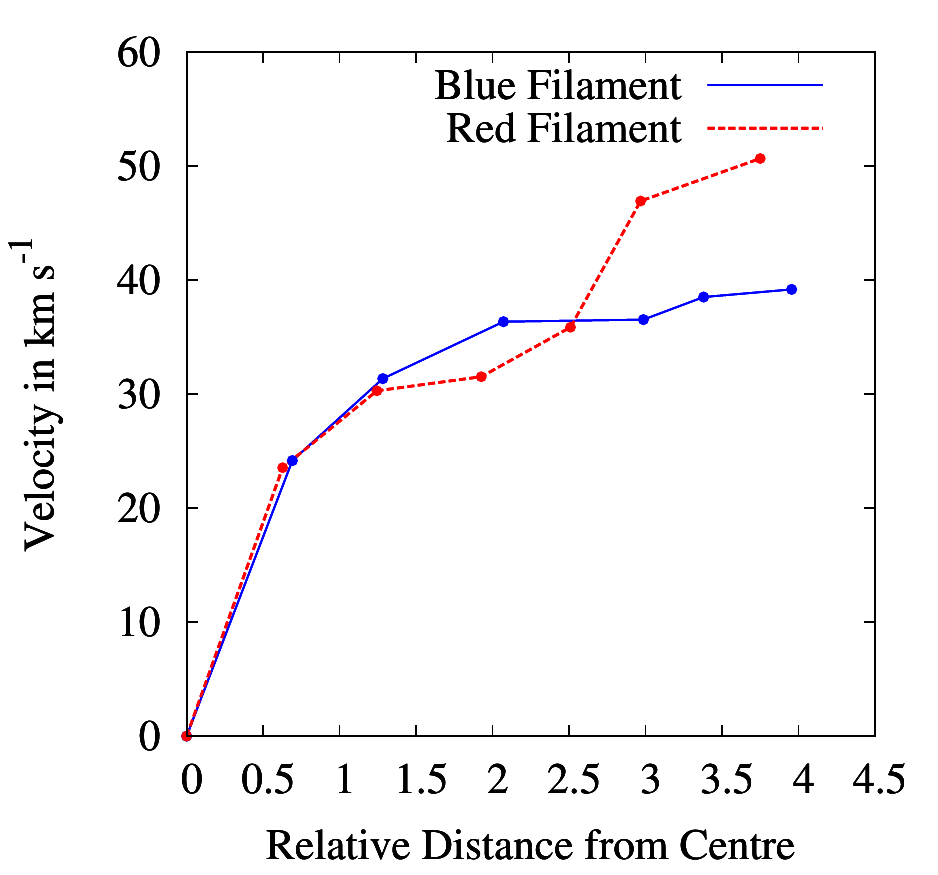}}
\caption{The velocity field in the redward (red dotted line) and blueward (blue solid line) filament arms of PC 19, obtained during fitting of the PV diagrams shown in Figure \ref{fig:pc19pv}(b). \label{fig:velprofile}}
\end{figure}

\begin{table*}
\centering
\caption{Parameters of 3D morphological model of PC 19. \label{tab:3Dparpc19}}
\small
\begin{tabular}{l c c c c c}
\hline
\hline
          &          & Major/ & Position            & Inclination         & Max. Expansion \\
Component & Geometry & Minor  & Angle               & Angle               & Velocity \\
          &          & Axis   & $(\alpha$ $^\circ)$ & $(\theta$ $^\circ)$ & (km s$^{-1}$) \\
\hline
Central Structure & Bipolar & 2.75 & 355 & 30 & 50 \\
Filament Pair & Bipolar and Spiral & - & - & $\sim$30 & 40 (blue); 52 (red) \\
\hline
\end{tabular}
\end{table*}

\subsubsection{Nebular Morphology: 3D Modeling} \label{sec:3dpc19}

The narrow band H$\alpha$, [O~{\sc iii}] and [N~{\sc ii}] images (Figure \ref{fig:pc19iac}) of PC 19 taken from IAC Catalog provide clues on morphological features collectively. Both the H$\alpha$ and [O~{\sc iii}] images depict a bright central region, whereas, the [N~{\sc ii}] image clearly shows a pair of filaments with a faint central region. The filaments had been characterized as low ionization regions \citep{1997IAUS..180..197L, 1999AJ....117..967G, 2001ApJ...547..302G}, i.e., they are more bright in low ionization lines, such as [N~{\sc ii}]. \citet{1999AJ....117..967G} studied the [N~{\sc ii}] (IAC Catalog image) morphology and kinematics of PC 19 and described the filaments as point symmetric arcs embedded in an elliptical shell. They also pointed out a velocity field increasing outwards along the filaments, with outflow velocities $\sim$20-25 km s$^{-1}$ and estimated a heliocentric velocity of 10 km s$^{-1}$. \cite{2001ApJ...547..302G} suggested the filaments to be outside the central nebula with higher velocities than the environment. The high-resolution HST image in [N~{\sc ii}] (Figure \ref{fig:3Dmodelpc19}(a)) clearly resolves the filament pair spiralling outward. A faint diffused central structure is also seen, but without lobes or shells and the filaments arise at a finite radius from the centre (SMV11). 

The 3D morphology of PC 19 is constructed mainly based on high-resolution [N~{\sc ii}] image (Figure \ref{fig:3Dmodelpc19}(a)). We model the filament pair by constructing highly collimated structure in SHAPE by applying \textit{size} and \textit{squeeze} modifiers on a basic \textit{sphere}. Then we obtain its point symmetric and spiral appearance using \textit{spiral} and \textit{twist} modifiers together. Constant volume-density distribution of \textit{particles} is used to replicate the clumpy effect of the arms. However, we faced difficulty to model the central region of PC 19 using the [N~{\sc ii}] image only, as, no high-resolution images comparable to HST images are available for PC 19 in higher-ionization lines (H$\alpha$/[O~{\sc iii}]) to resolve the central structure effectively. Hence, we take the reference of the PV diagram in H$\alpha$ and [N~{\sc ii}] from \cite{SPMCatalog} (Figure \ref{fig:pc19pv}(a)). In the [N~{\sc ii}] PV profile, the two blobs are seen, which correspond to the positions and velocities of the filament pair having velocities blueward and redward, with no prominent evidence of the central portion. Whereas, in H$\alpha$ PV profile, the central structure is prominent. Hence, combining the available clues from the PV diagrams and IAC images, we assume a bipolar central structure with partially open lobes. Along with the mesh, the structure is modelled with distribution of \textit{particles} in such a way that the density would, in effect, appear decreasing along the major axis and increasing radially outward along the direction of the minor axis. A velocity field proportional to the distance from the central star is assumed. 

As discussed earlier also, simultaneous estimation of the expansion velocity, inclination angle and length of the component involves a certain amount of degeneracy, thus, making difficult to obtain a unique set of parameters. After trying different combination of these parameters, we obtain an optimized set for these parameters from the model that fits best according to eye-estimation. 

We study the velocity field of the filaments using [N~{\sc ii}] spectra (PV diagrams) (Figure \ref{fig:pc19pv}(b)) taken from Guerrero et al. (1999). The spectra were obtained at position angles 0$^\circ$ and 28$^\circ$. The different slit orientations cover two different regions in each filament, giving PV correlation at two regions along the length of each filament. 
Using this information, we aim to construct the radial velocity structure (Figure \ref{fig:velprofile}) of the outflowing matter in the filaments. We use the custom velocity profile in the \textit{velocity} modifier defining the velocity field. We adjust the velocity values along the radius to match the observed PV diagrams for both the slit orientations. Finally, we obtain best-fitting result of the profile by eye-estimation. Also, we obtain an estimation of the maximum expansion velocity of the central region as 50 km s$^{-1}$, by matching the faint structures at the central part of the PV diagrams. The structure is inclined at an angle of 30$^\circ$ with the line of sight creating a position angle of $355^\circ$. The major-to-minor axis ratio for the central structure is found to be 2.75. The filaments are approximately aligned with the major axis of the central structure. Simultaneous estimation of the expansion velocity, inclination angle and length of the component, involves the problem of degeneracy. 

The parameters of the 3D model are summarized in Table \ref{tab:3Dparpc19}. In Figure \ref{fig:3Dmodelpc19} the observed [N~{\sc ii}] image and the rendered synthetic image (both in grey-scale) are shown side-by-side, depicting the 2D morphology being well-reproduced by our model. The side view, top view, front view and sky view from the Earth of the 3D mesh model of nebular structure are also shown. 

\begin{table}
\centering
\small
\caption{Temperatures and densities for PC 19. \label{tab:specanalysispc19}}
 \begin{tabular}{lccc}
 \hline
 & This Work & C96$^a$ & O12$^b$\\
 \hline
$N_\mathrm{e}$([Ar~{\sc iv}]) & ${5720}$ & - & -\\
$N_\mathrm{e}$([Cl~{\sc iii}]) & ${3694}$ & - & -\\
$N_\mathrm{e}$([S~{\sc ii}]) & ${4790}$ & ${9510}$ & ${14542\pm5383}$\\
$T_\mathrm{e}$([O~{\sc iii}]) & ${12548}$ & ${11400}$ & ${13472\pm211}$\\
$T_\mathrm{e}$([N~{\sc ii}]) & ${15183}$ & - & ${10570\pm1278}$\\
$T_\mathrm{e}$([S~{\sc iii}]) & ${14466}$ & - & -\\
 \hline
\multicolumn{2}{l}{$^a$Reference: \citet{1996A&A...307..215C}}\\
\multicolumn{2}{l}{$^b$Reference: \citet{2012IAUS..283..464O}}\\
 \end{tabular}
\end{table}

\begin{figure*}
\centering
\scalebox{0.685}[0.685]{\includegraphics{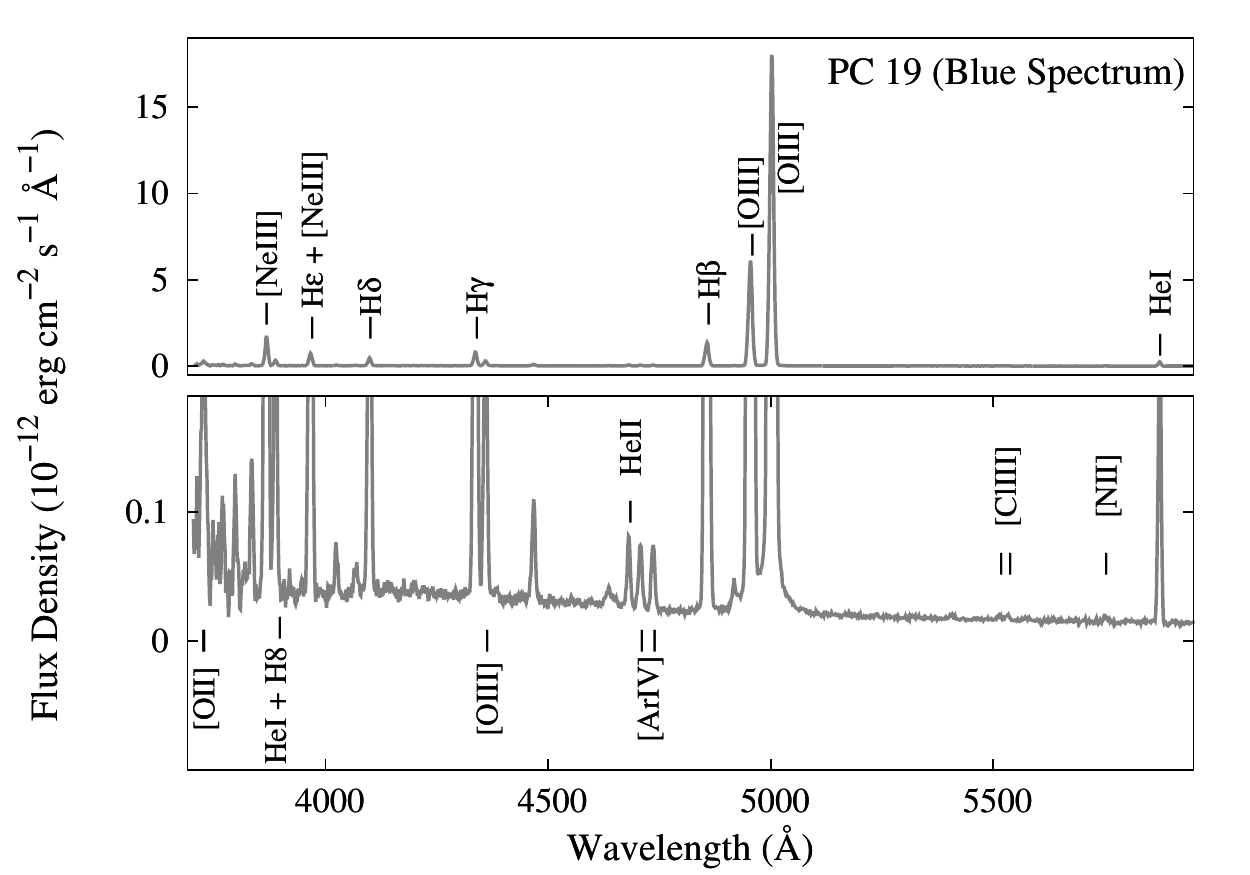}}
\scalebox{0.685}[0.685]{\includegraphics{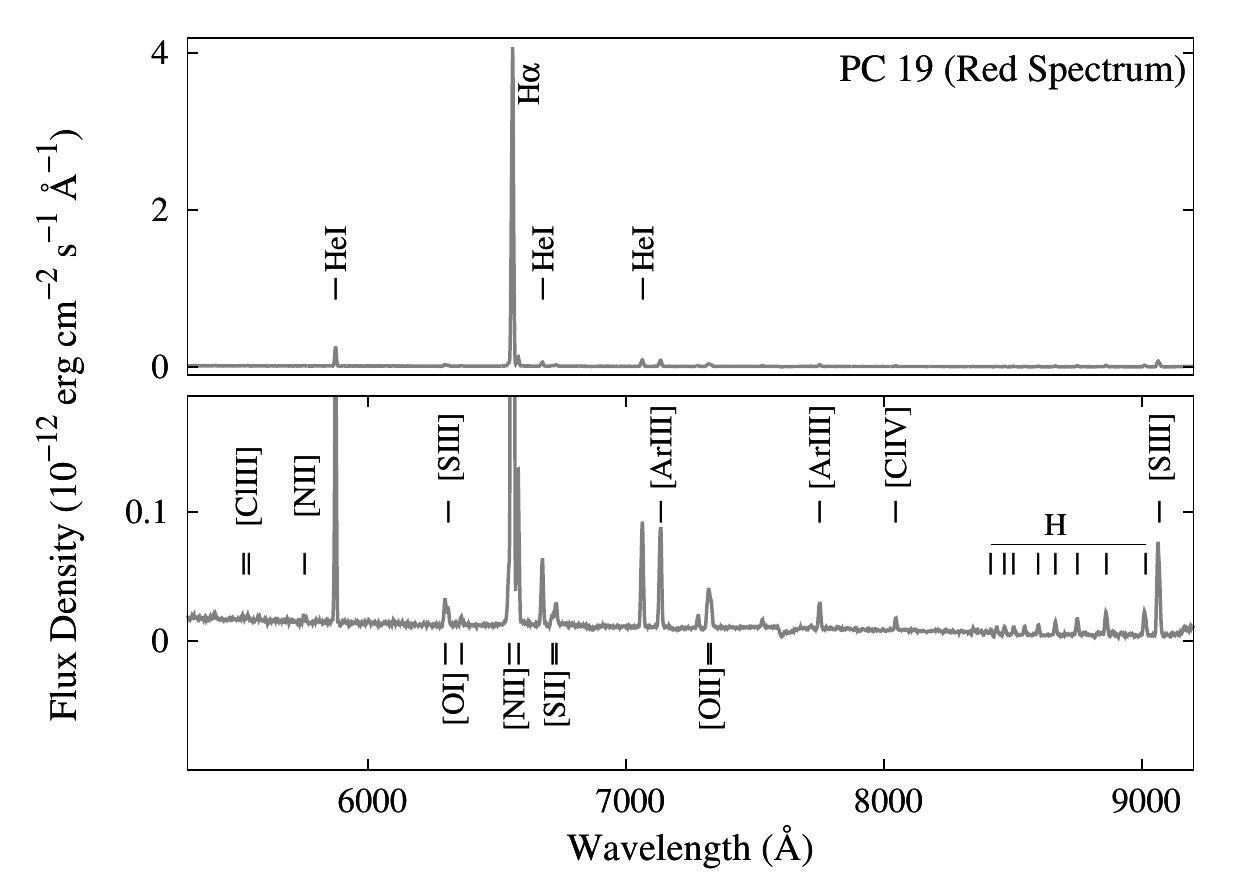}}
 \caption{The reduced dereddened blue (left panel) and red spectrum (right panel) of PC 19 obtained through Gr. 7 and Gr. 8, respectively. The spectra are zoomed vertically near the continuum to make the weaker lines visible (lower panel in both the images). Prominent emission lines are marked. Fluxes are in absolute scale. \label{fig:obsspecpc19}}
\end{figure*}

\begin{figure}
\centering
\scalebox{0.55}[0.55]{\includegraphics{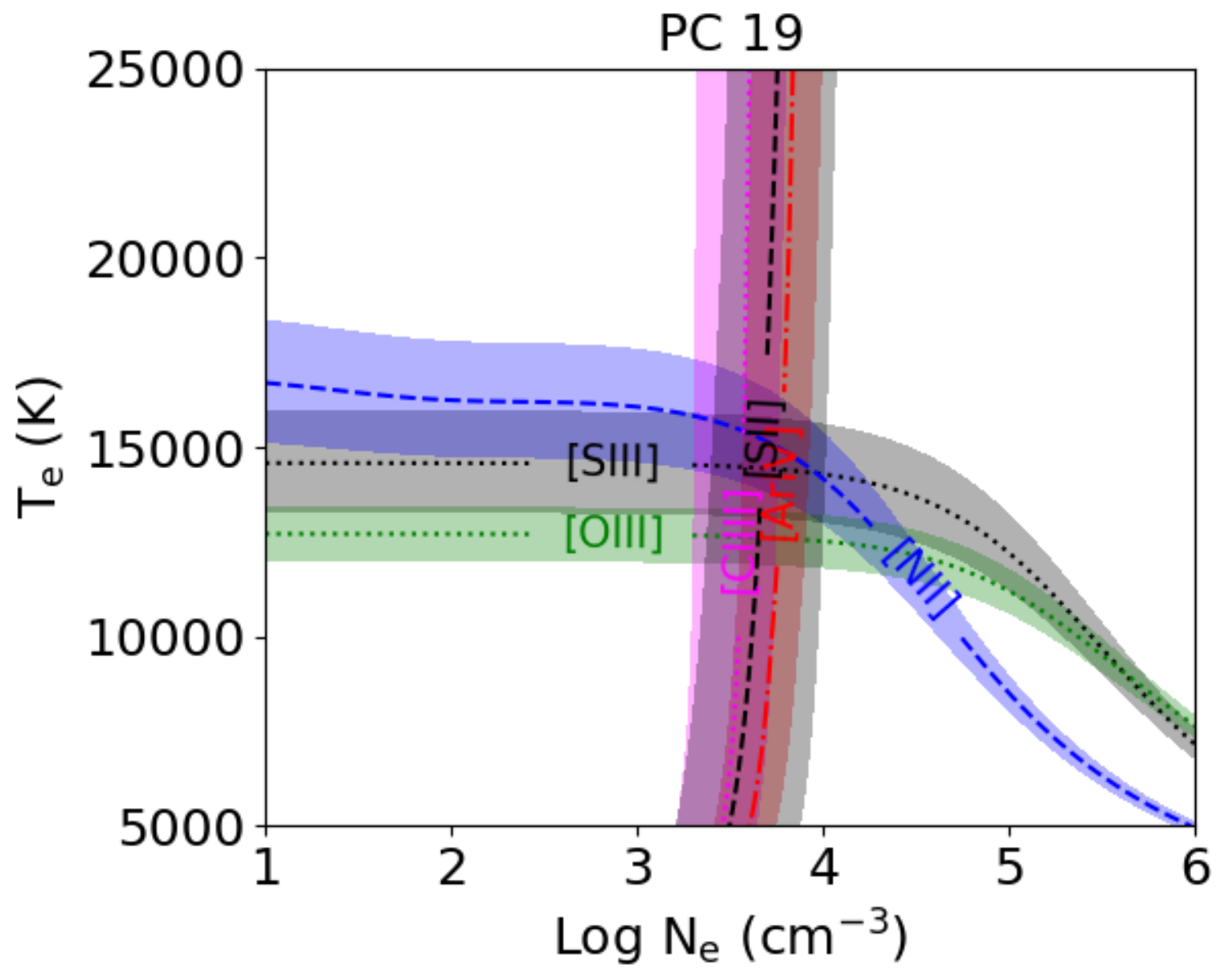}}
 \caption{The $T_\mathrm{e}$ vs Log $N_\mathrm{e}$ plot for PC 19. \label{fig:TeNepc19}}
\end{figure}

\subsubsection{Analysis of Optical Spectrum}
\subsubsection*{Emission Lines and Fluxes}
The blue and red spectrum of PC 19 are separately shown in Figure \ref{fig:obsspecpc19} and the prominent lines are marked. The spectrum of the two regions are combined to get a single spectrum as done for PB 1.
The RLs of hydrogen (H$\alpha$ 6563, H$\beta$ 4861, H$\gamma$ 4340, H$\delta$ 4101,and  H$\epsilon$ 3970 {\AA}) and helium (He~{\sc i} 4471, 5876, 6678, and 7065 {\AA}) are prominently seen in the spectrum. Among the He~{\sc ii} lines, only He~{\sc ii} 4686 {\AA} is prominent. Among the CELs, we detect strong [O~{\sc iii}] 4959, 5007 {\AA} and the auroral [O~{\sc iii}] 4363 {\AA} lines. [N~{\sc ii}] 6548, 6583 {\AA} lines are stronger compared to PB 1 and the auroral [N~{\sc ii}] 5755 {\AA} is also detected. Other prominently detected lines are [Ne~{\sc iii}] 3869 {\AA} ,[Ar~{\sc iii}] 7136, 7751 {\AA}, [S~{\sc iii}] 6312, 9069 {\AA}, [Cl~{\sc iv}] 7531, 8046 {\AA}, [O~{\sc i}] 6300 (partially blended with [S~{\sc iii}] 6312 {\AA}), 6364 {\AA}, [Ar~{\sc iv}] 4711, 4740 {\AA}, [S~{\sc ii}] 6716, 6731 {\AA} and [Cl~{\sc iii}] 5518, 5538 {\AA}.
We obtain $c(\mathrm{H}\beta)={1.29}$, in order to match the theoretical Balmer line ratio, H$\alpha$/H$\beta$, in the dereddened spectrum. We measured the H$\beta$ flux through slit to be, $F(\mathrm{H}\beta)=2.64\times10^{-13}$ erg cm$^{-2}$ s$^{-1}$.

\subsubsection*{Plasma Diagnostics} \label{sec:pldiagpc19}

The physical conditions inside the nebulae are analysed using PyNeb package. We calculate [O~{\sc iii}], [N~{\sc ii}], [S~{\sc iii}] electron temperatures ($T_\mathrm{e}$) and [Ar~{\sc iv}], [Cl~{\sc iii}], [S~{\sc ii}] electron densities ($N_\mathrm{e}$). The results are shown in Table \ref{tab:specanalysispc19}. We list the electron temperatures and densities estimated by \citet{1996A&A...307..215C} and \citet{2012IAUS..283..464O} for comparison. The contribution of He~{\sc i} 4713 {\AA} from [Ar~{\sc iv}] 4711 {\AA} has been subtracted before analysis, as done in case of PB 1 (Sec. \ref{sec:pldiagpb1}). Figure \ref{fig:TeNepc19} shows the $T_\mathrm{e}$-$N_\mathrm{e}$ map for the ions mentioned above. We observe variations among the temperatures for the three different ions. For the calculation of ionic abundances, we adopt the values, $T_e\sim13500$ K and $n_e\sim4700$ cm$^{-3}$. Using the direct method, the total elemental abundances from CELs of N, O, Ne, Ar, S and Cl are calculated adopting the ICF expressions from \citet{2014MNRAS.440..536D}. The calculated ionic abundances, total elemental abundances and the ICF values are given in Table \ref{tab:modelparpc19}. C/H cannot be calculated as no C lines are detected in our spectrum.  

\subsubsection{Ionization Characteristics: Photoionization Modeling} \label{sec:pimodelpc19}

\begin{table}
\centering
\footnotesize
\caption{Results of photoionization modelling for PC 19. \label{tab:modelparpc19}}
\begin{tabular}{l c c}
%\hline
\hline
r.m.s. && 0.21\\
H$\beta$ Flux (Total) & Log $I(\mathrm{H}\beta)$ & -10.79\\
H$\beta$ Flux (Low-den. Comp.) & Log $I(\mathrm{H}\beta)$ & -10.91\\
H$\beta$ Flux (High-den. Comp.) & Log $I(\mathrm{H}\beta)$ & -11.39\\
Distance to PN & $d$ $(\mathrm{kpc})$ & 5.57\\
&&\\
\multicolumn{3}{l}{Central Star}\\
Temperature & $T_\mathrm{eff}$ $(\mathrm{K})$ & 92000\\
Luminosity & $L$ $(L_\mathrm{\sun})$ & 8309\\
Gravity & Log $g$ $(\mathrm{cm}$ $\mathrm{s^{-2}})$ & 5.8\\
&&\\
\multicolumn{3}{l}{Nebula: low-den. comp. (Central Structure)}\\
Geometry && Closed\\
Hydrogen Density & $n_\mathrm{H}$ $(\mathrm{cm^{-3}})$ & 3.6\\
Inner Radii & $r_\mathrm{in}$ $(\mathrm{pc})$ & 0.029\\
Outer Radii & $r_\mathrm{out}$ $(\mathrm{pc})$ & 0.081\\
Filling Factor & $f$ & 0.45\\
&&\\
\multicolumn{3}{l}{Nebula: high-den. comp. (Filaments)}\\
Geometry && Open\\
Hydrogen Density & $n_\mathrm{H}$ $(\mathrm{cm^{-3}})$ & 4.0\\
Inner Radii & $r_\mathrm{in}$ $(\mathrm{pc})$ & 0.029\\
Outer Radii & $r_\mathrm{out}$ $(\mathrm{pc})$ & 0.091\\
Filling Factor & $f$ & 0.9\\
Covering Factor & $\mathrm{\Omega}$ & 0.01\\
\hline
\hline
\end{tabular}
\begin{tabular}{l c c c c c}
\multicolumn{6}{l}{Total Elemental Abundances}\\
& Model & DM$^a$ & C96$^b$ & S06$^c$ & O12$^d$\\
\hline
He/H & 0.094 & 0.086 & 0.11 & 0.113 & 0.103\\
C/H$\times10^{4}$ & 2.69 & - & - & - & -\\
N/H$\times10^{4}$ & 0.47 & 0.42 & 0.32 & 0.37 & 0.378\\
O/H$\times10^{4}$ & 2.5 & 2.01 & 3.6 & 3.09 & 4.292\\
Ne/H$\times10^{4}$ & 0.59 & 0.38 & - & - & -\\
S/H$\times10^{5}$ & 0.27 & 0.32 & - & - & -\\
Cl/H$\times10^{7}$ & 0.63 & 0.44 & 0.347 & - & 0.461\\
Ar/H$\times10^{6}$ & 0.85 & 0.47 & 0.97 & - & 1.279\\
\hline
\hline
\end{tabular}
\begin{tabular}{l c c c c}
\multicolumn{5}{l}{Ionic Abundances using Direct Method}\\
& X$^{+}$/H & X$^{2+}$/H & X$^{3+}$/H & ICF\\
\hline
He	&	0.0824	&	0.0033	&	-		& 1.0\\
C	&	-		&	-		&	-   		& -\\
N	&	1.21(-6)	&	-		&	-		& 33.82\\
O	&	6.86(-6)	&	1.94(-4)	&	-	 	& 1.02\\
Ne	&	-		&	3.71(-5)	&	-		& 1.03\\
S	&	5.54(-8)	&	6.68(-7)	&	-		& 56.91\\
Cl	&	-		&	2.06(-8)	&	2.48(-8)	& 2.1\\
Ar	&	-		&	2.81(-7)	&	4.32(-7)	& 1.62\\
\hline
\hline
\end{tabular}
\begin{tabular}{l c c c c}
\multicolumn{5}{l}{Model Ionization Fractions weighted by $N_\mathrm{e}$}\\
& X$^{+}$ & X$^{2+}$ & X$^{3+}$  & ICF\\
\hline
He	&	0.9584	&	0.0389	&	-	&	1.0\\
C	&	0.0203	&	0.5200	&	0.4573	&	1.92	\\
N	&	0.0307	&	0.4560	&	0.5125	&	32.59	\\
O	&	0.0276	&	0.9438	&	0.0223	&	1.03	\\
Ne	&	0.0086	&	0.9772	&	0.0142	&	1.02	\\
S	&	0.0239	&	0.3026	&	0.6060	&	41.86	\\
Cl	&	0.0166	&	0.2911	&	0.6859	&	3.43	\\
Ar	&	0.0042	&	0.3275	&	0.6632	&	3.05	\\
\hline
\multicolumn{5}{l}{$^a$DM stands for `Direct Method'}\\
\multicolumn{5}{l}{$^b$Reference: \citet{1996A&A...307..215C}}\\
\multicolumn{5}{l}{$^c$Reference: \citet{2006ApJ...651..898S}}\\
\multicolumn{5}{l}{$^d$Reference: \citet{2012IAUS..283..464O}}\\
\end{tabular}
\end{table}

\begin{table*}
\centering
\small
\caption{Comparison between observed and modelled line intensities of PC 19. Observed and modelled flux values are given with respect to $I(\mathrm{H}\beta)=100$. The fifth and sixth column show the contribution of the low-density component (central structure) and the high-density component (filaments), respectively. Emission line fluxes are normalized to $I(\mathrm{H}\beta)=75$ and $I(\mathrm{H}\beta)=25$ for the low- and high-density components, respectively (see Sec. \ref{sec:pimodelpc19} for details). \label{tab:obsvsmodpc19}}
%\small
\begin{tabular}{c c c c c c c c}
%\hline
\hline
&& \multicolumn{2}{c}{Flux} & \multicolumn{2}{c}{Modeled Flux Contribution} &&\\
Wavelength ({\AA}) & Line Id. & Observed & Modeled & Low-den. Comp. & High-den. Comp. & Mod./Obs. & $\kappa_\mathrm{i}$\\
\hline
3727	&	[O~{\sc	ii}]				&	21.43	&	22.35	&		4.17		&		18.18	&	1.04	&	0.44	\\
3869	&	[Ne~{\sc	iii}]				&	101.47	&	101.16	&		72.22	&		28.94	&	1.00	&	-0.03	\\
3889	&	He~{\sc	i}	,	H8		&	18.47	&	18.11	&		14.59	&		3.52		&	0.98	&	-0.21	\\
3968	&	[Ne~{\sc	iii}]	,	H$\epsilon$		&	47.21	&	46.79	&		34.05	&		12.74	&	0.99	&	-0.09	\\																							
4026	&	He~{\sc	ii}	,	He~{\sc	i}	&	2.54	&	2.21	&		1.62		&		0.59		&	0.87	&	-0.77	\\
4101	&	H$\delta$					&	27.90	&	26.18	&		19.74	&		6.44		&	0.94	&	-0.67	\\
4340	&	H$\gamma$					&	51.55	&	47.38	&		35.54		&		11.84		&	0.92	&	-0.88	\\
4363	&	[O~{\sc	iii}]				&	18.43	&	18.51	&		12.97		&		5.54		&	1.00	&	0.05	\\
4388	&	He~{\sc	i}				&	0.74	&	0.58	&		0.43		&		0.15		&	0.78	&	-0.92	\\
4471	&	He~{\sc	i}				&	5.33	&	4.75	&		3.46		&		1.29		&	0.89	&	-0.63	\\
4686	&	He~{\sc	ii}				&	3.80	&	3.29	&		3.10		&		0.18		&	0.87	&	-0.79	\\
4711	&	[Ar~{\sc	iv}]				&	3.18	&	4.44	&		3.71		&		0.73		&	1.40	&	1.83	\\
4740	&	[Ar~{\sc	iv}]				&	3.88	&	3.56	&		2.91		&		0.65		&	0.92	&	-0.47	\\
4861	&	H$\beta$					&	100.00	&	100.00	&		75.00		&		25.00		&	1.00	&	0.00	\\
4959	&	[O~{\sc	iii}]				&	466.06	&	468.24	&		345.73		&		122.52		&	1.00	&	0.05	\\
5007	&	[O~{\sc	iii}]				&	1402.71	&	1397.08	&		1031.55		&		365.53		&	1.00	&	-0.04	\\
5411	&	He~{\sc	ii}				&	0.28	&	0.28	&		0.26		&		0.02		&	1.00	&	-0.01	\\
5518	&	[Cl~{\sc	iii}]				&	0.33	&	0.20	&		0.13		&		0.07		&	0.60	&	-1.96	\\
5538	&	[Cl~{\sc	iii}]				&	0.36	&	0.30	&		0.17		&		0.14		&	0.84	&	-0.65	\\
5755	&	[N~{\sc	ii}]				&	0.45	&	0.36	&		0.02		&		0.35		&	0.81	&	-0.81	\\
5876	&	He~{\sc	i}				&	14.34	&	14.16	&		10.12		&		4.04		&	0.99	&	-0.13	\\
6300	&	[O~{\sc	I}]				&	1.57	&	1.28	&		0.00		&		1.28		&	0.82	&	-1.12	\\
6312	&	[S~{\sc	iii}]				&	0.96	&	0.83	&		0.46		&		0.38		&	0.87	&	-0.52	\\
6364	&	[O~{\sc	I}]				&	0.50	&	0.41	&		0.00		&		0.41		&	0.81	&	-0.80	\\
6548	&	[N~{\sc	ii}]				&	4.00	&	3.94	&		0.21		&		3.73		&	0.98	&	-0.09	\\
6563	&	H$\alpha$					&	283.56	&	276.65	&		206.55		&		70.10		&	0.98	&	-0.26	\\
6583	&	[N~{\sc	ii}]				&	11.53	&	11.61	&		0.61		&		11.00		&	1.01	&	0.07	\\
6678	&	He~{\sc	i}				&	3.84	&	3.49	&		2.54		&		0.95		&	0.91	&	-0.53	\\
6716	&	[S~{\sc	ii}]				&	0.78	&	0.53	&		0.07		&		0.45		&	0.67	&	-1.51	\\
6731	&	[S~{\sc	ii}]				&	1.38	&	0.86	&		0.11		&		0.75		&	0.62	&	-2.61	\\
7065	&	He~{\sc	i}				&	6.47	&	9.13	&		5.75		&		3.38		&	1.41	&	1.88	\\
7136	&	[Ar~{\sc	iii}]				&	6.44	&	5.21	&		2.82		&		2.39		&	0.81	&	-1.17	\\
7281	&	He~{\sc	i}				&	0.76	&	0.92	&		0.65		&		0.27		&	1.21	&	0.73	\\
7325	&	[O~{\sc	ii}]				&	4.30	&	4.42	&		0.52		&		3.91		&	1.03	&	0.16	\\
7531	&	[Cl~{\sc	iv}]				&	0.48	&	0.38	&		0.31		&		0.07		&	0.80	&	-0.87	\\
7751	&	[Ar~{\sc	iii}]				&	1.69	&	1.24	&		0.67		&		0.57		&	0.73	&	-1.71	\\
8046	&	[Cl~{\sc	iv}]				&	0.65	&	0.89	&		0.72		&		0.16		&	1.36	&	1.17	\\
8323	&	P25					&	-	&	1.13	&		0.89		&		0.24		&	-	&		\\
8334	&	P24					&	-	&	0.45	&		0.32		&		0.13		&	-	&		\\
8346	&	P23					&	0.27	&	0.31	&		0.22		&		0.09		&	1.16	&	0.58	\\
8359	&	P22					&	0.24	&	0.28	&		0.20		&		0.08		&	1.19	&	0.66	\\
8374	&	P21					&	0.22	&	0.29	&		0.21		&		0.07		&	1.30	&	1.01	\\
8392	&	P20					&	0.16	&	0.31	&		0.23		&		0.08		&	1.97	&	2.58	\\
8413	&	P19					&	0.27	&	0.35	&		0.26		&		0.09		&	1.28	&	0.94	\\
8438	&	P18					&	0.58	&	0.40	&		0.30		&		0.10		&	0.68	&	-1.47	\\
8467	&	P17					&	0.51	&	0.46	&		0.35		&		0.11		&	0.90	&	-0.41	\\
8502	&	P16					&	0.51	&	0.54	&		0.41		&		0.13		&	1.07	&	0.27	\\
8545	&	P15					&	0.48	&	0.65	&		0.49		&		0.16		&	1.36	&	1.17	\\
8598	&	P14					&	0.63	&	0.79	&		0.60		&		0.19		&	1.24	&	0.83	\\
8665	&	P13					&	0.81	&	0.97	&		0.73		&		0.24		&	1.20	&	0.69	\\
8750	&	P12					&	1.01	&	1.21	&		0.92		&		0.30		&	1.21	&	1.03	\\
8863	&	P11					&	1.54	&	1.55	&		1.17		&		0.38		&	1.01	&	0.04	\\
9015	&	P10					&	1.80	&	1.78	&		1.35		&		0.44		&	0.99	&	-0.03	\\
9069	&	[S~{\sc	iii}]				&	7.00	&	8.20	&		4.79		&		3.40		&	1.17	&	0.87	\\
\hline
\end{tabular}
\end{table*}

\begin{figure*}
\centering
\scalebox{0.9}[0.9]{\includegraphics{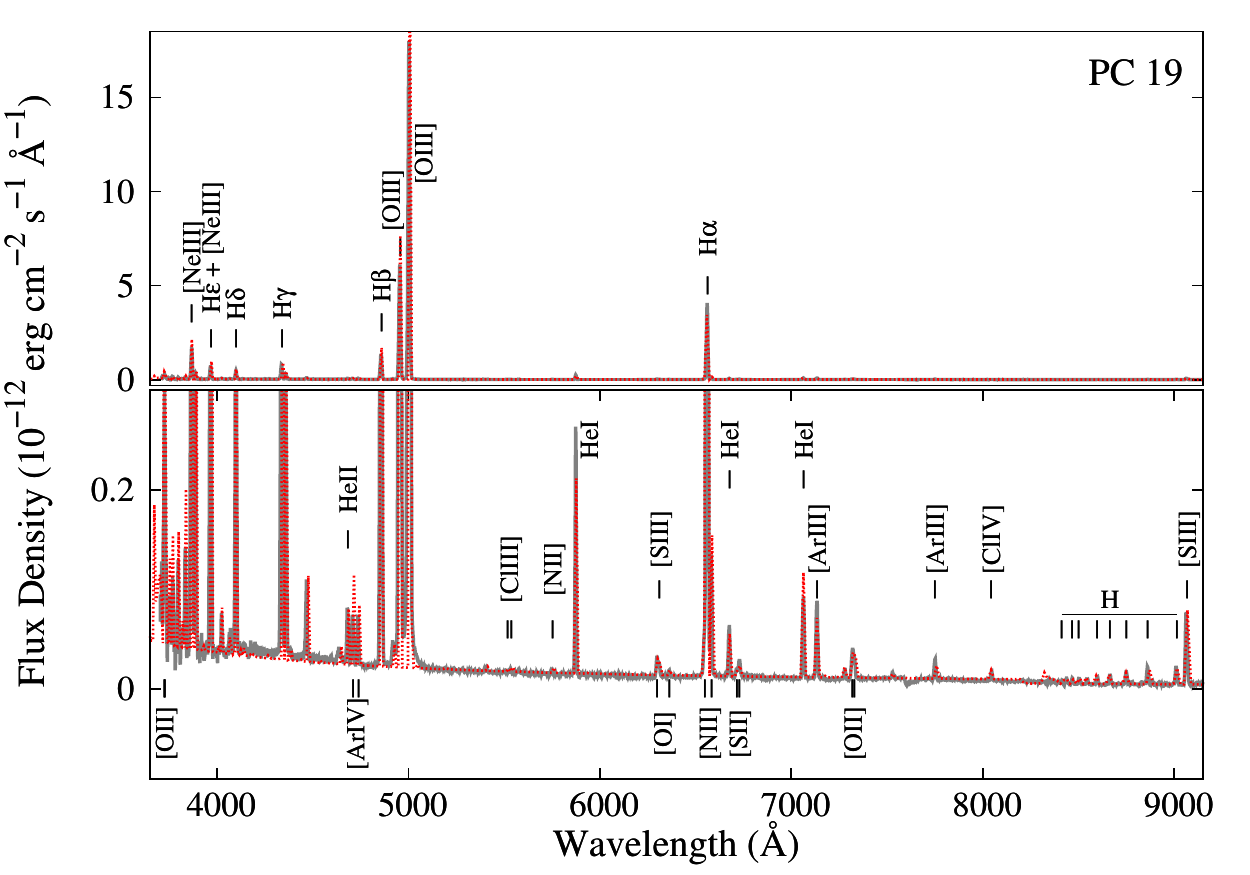}}
\caption{The observed optical spectrum (grey solid line) and the modelled spectrum (red dashed line) are shown for PC 19. Prominent emission lines are marked. Fluxes are in absolute scale. The vertically zoomed spectrum (lower panel) shows the fit of the weaker lines and the continuum clearly. See Section \ref{sec:pimodelpc19} for details. \label{fig:pimodelpc19}}
\end{figure*}

The photoionization model of PC 19 is difficult to construct with simple assumption of a spherical geometry due the presence of the filament pair as morphological elements. While we have tried to construct the model with one component spherical geometry, we have found that neither a high-density nor a low-density model could simultaneously reproduce all the characteristic observables of the PN, e.g., H$\beta$ flux, the continuum, and the characteristic lower to higher ionization line ratios. 
Thus, at first, to qualitatively obtain all the characteristic observables in our photoionization model, we attempt to construct a two-component model. A low-density component with closed spherical geometry is considered to represent the central structure. To represent the filaments, we consider a high-density component having open geometry and an appropriate covering factor. In each iteration, the parameters characterizing the central star (temperature, luminosity and gravity) and the nebular abundance values are kept same for both the components, while the nebular parameters, such as, hydrogen density, radii, filling factor, etc., are different. The output model spectra (in absolute fluxes) from the two components are directly summed to obtain the total model spectrum.  

From the 3D morphological model it could be seen that the filaments cover very low fraction of space around the central star. This guided us to use a very small covering factor for the high-density component. In a Cloudy model, for any change of covering factor of the ionized shell, the model absolute flux values vary linearly, but the line flux ratios in the output model spectrum (considering one-component model) do not change considerably. In case of a two-component model as we have considered for PC 19, due to any change of covering factor of the high-density component, the line flux ratios in the total model spectrum (sum of contributions from the low- and high-density components) also change. Thus, an independent estimation of the covering factor from the morphology helps to put constrain on covering factor in the Cloudy model. From an approximate calculation of the radial extent and average cross-section of the filaments, we find that the covering factor of the high-density component should be $\sim$1$\%$ or $\sim$0.01. We varied the covering factor slightly around 0.01 and compared the model spectrum with the observed spectrum. Finally, from visual inspection, we adopt 0.01 as the value of covering factor of the high-density component in our model. 

Figure \ref{fig:pimodelpc19} shows the fit of the observed spectrum with the model spectrum. We followed the similar fitting procedure as discussed in case of PB 1 (Sec. \ref{sec:pimodelpb1}) to obtain a satisfactory model. The comparison between the observed and modelled fluxes of the emission lines are given in Table \ref{tab:obsvsmodpc19}. The lines particularly focussed on for fitting the photoionization model include He~{\sc i} 5876 {\AA}, He~{\sc ii} 4686 {\AA}, [O~{\sc i}] 6364 {\AA}, [O~{\sc ii}] 3727 {\AA}, [O~{\sc iii}] 5007, 4363 {\AA}, [N~{\sc ii}] 6583, 5755 {\AA}, [Ne~{\sc iii}] 3869 {\AA}, [S~{\sc ii}] 6716, 6731, [S~{\sc iii}] 9069 {\AA}, [Ar~{\sc iv}] 4711, 4740 {\AA}, [Ar~{\sc iii}] 7136 {\AA}, [Cl~{\sc iii}] 5518, 5538 {\AA}, [Cl~{\sc iv}] 8045 {\AA}. We also aim to carefully match the plasma diagnostic line ratios, [O~{\sc iii}] 5007/4363, [N~{\sc ii}] 6583/5755, [S~{\sc iii}] 9069/6312, [S~{\sc ii}] 6731/6716, [Ar~{\sc iv}] 4740/4711 and [Cl~{\sc iii}] 5538/5518. However, the ratios, [S~{\sc iii}] 9069/6312, [Ar~{\sc iv}] 4740/4711 and [Cl~{\sc iii}] 5538/5518 are not reproduced well by our model. The lines of [O~{\sc i}] and [S~{\sc ii}] lines are slightly underestimated in our model.

The fifth and sixth columns of Table \ref{tab:obsvsmodpc19} show the contribution of the low-density component (central structure) and the high-density component (filaments), respectively, to the total model emission line fluxes. From best-fitting model, we find that the low-density component and the high-density component contribute 75$\%$ and 25$\%$ of the total model $\mathrm{H}\beta$ flux, respectively. 

We obtain a satisfactory match of the modelled continuum and the absolute line fluxes with their observed values. The absolute H$\beta$ flux, Log $I(\mathrm{H}\beta)=-10.79$, obtained from model, matches satisfactorily with the observed value of Log $I(\mathrm{H}\beta)=-10.91$. The goodness of fit of the model is obtained by calculating r.m.s value of the fit. We calculate a value of 0.21, within the considerable range. The quality of fit of individual emission lines, measured with a quality factor ($\kappa_\mathrm{i}$), as described earlier, is given in Table \ref{tab:obsvsmodpc19}. 

From the best-fitting model, we find effective temperature of the central star to be $92000$ K, close to that estimated by \citet{2002ApJ...576..285S} ($T_\mathrm{eff}=88920$ K). Central star gravity is estimated as Log $g=5.8$ cm s$^{-2}$. The low- and high-density components are estimated to have densities of $\sim$4000 cm$^{-1}$ and $\sim$10000 cm$^{-3}$, respectively. Similar to PB 1, due to the presence of filling factor in the model, the simultaneous measurements of distance (and radii), luminosity and filling factor may involve degeneracy. The estimation of these parameters is described in Sec. \ref{sec:distpc19}. 

In Table \ref{tab:modelparpc19}, we present the abundance values estimated from photoionization modelling and those obtained using direct method (Sec. \ref{sec:pldiagpc19}). We show the abundance values obtained by \citet{1996A&A...307..215C} (hereafter C96), \citet{2006ApJ...651..898S} (hereafter S06) and \citet{2012IAUS..283..464O} (hereafter O12) for comparison. We also list the ionic abundances and ICF values calculated for obtaining the total abundances from direct method, and the model ionic fractions of the elements along with the corresponding ICF values in Table \ref{tab:modelparpc19}.

Our estimated abundances match well with other studies. However, few mismatches are there. For example, our estimated N/H from model and direct method (using ICF) are slightly higher than that obtained by C96, S06 and O12. In case of O/H, our values (from model and direct method) are slightly lower than the other estimations. The Ar/H obtained using ICF in our analysis is about half compared to our model value and also to the other estimations. Our model value of Ne/H is $\sim$1.5 times higher than our estimated value using ICF. Cl/H obtained from our model is slightly higher than our estimated value using ICF and also than the other estimations. We could not constrain C/H for PC 19 by photoionization modelling since no C lines are present in our spectrum. We kept the C abundance constant at solar value \citep{2010Ap&SS.328..179G} throughout the modelling. We find no previous estimation of C/H, Ne/H and S/H to compare with our results.    

While computing the photoionization model, we consider few probable aspects to verify the feasibility of the photoionization model due to the presence of low ionization filaments. We discuss this below. The low ionization lines such as [N~{\sc ii}], are generally much stronger in most of the PNe with low ionization structures (LISs). Though abundance enhancement was proposed as a reason for this (e.g., \citealt{1993ApJ...411..778B}), more recently it was showed that the enhancement of low ionization lines in LISs are not necessarily due to abundance enhancement, but might be predominantly due to shock excitation (e.g., \citealt{2009MNRAS.398.2166G}; \citealt{2016MNRAS.455..930A}). The latter also showed that the regions of LISs have similar or lower electron densities as the surrounding nebula, whereas, theoretical models (e.g., \citealt{1997ApJ...485L..41D}; \citealt{2001ApJ...556..823S}) predict the density in the LISs to be higher than the rest of the nebula. The justification was given that while the theoretical model densities represent the total densities from atomic, ionic, dust and molecular contribution, the electron densities only correspond to ionized fractions. Although LISs are present, the low ionization lines (e.g., [N~{\sc ii}]) are not particularly strong in PC 19 as compared to many other PNe with such characteristics (e.g., \citealt{1999AJ....117..967G}). However, \citet{1999AJ....117..967G} showed that [N~{\sc ii}]/H$\alpha$ ratio increases by 11.23 in the filaments than in the central region. However, no abundance enhancement of N was noticed for this PN in any previous N/H estimation, viz. C96; S06; O12 (mentioned earlier), where the abundances of nitrogen were similar to that estimated in this work using direct method (Sec. \ref{sec:pldiagpc19}). We also inspected whether PC 19 seem to fall in the category where photoionization dominates the shock excitation. Presumably, the more shock excitation would dominate photoionization, when the velocity is higher, temperature $\&$ luminosity of the ionizing source is lower and the region under consideration is nearer the central star. We took the set of models computed by \citet{2008A&A...489.1141R} for a cloudlet moving with a velocity, $v_c=100$ km s$^{-1}$. The model with highest ionization (C100) was computed with the parameters, central source temperature, $T_\mathrm{eff}=7\times10^4$; luminosity, $L=5000 L_{\sun}$; distance of the nebula from the ionizing source, $D=3\times10^{17}$ cm. Comparing these values with our model parameters, i.e., temperature, luminosity and inner radii of  both the nebular components, it is evident that the model would fall in the category where photoionization dominates shock excitation.

\subsubsection{Distance, Nebular Radii and Luminosity} \label{sec:distpc19}
In the previous studies, distance to this PC 19 had been estimated in a wide range; e.g., $5.715$ \citep{1992A&AS...94..399C}, $6.89$ \citep{1995ApJS...98..659Z}, $2.55$ \citep{2002ApJS..139..199P}, $6.75$ \citep{2004MNRAS.353..589P}, $4.4$ \citep{2006ApJ...651..898S}, $10.390$ \citep{2008ApJ...689..194S}, and $4.002\pm0.8$ kpc \citep{2010ApJ...714.1096S}. To summarize, the distance has been estimated in a range of 2.55 to 10.39 kpc.

Similar modelling approach, as described in case of PB 1 (Sec. \ref{sec:distpb1}), is followed in order to estimate distance, radii, luminosity and filling factor. However, unlike PB 1, PC 19 has complex morphology comprising highly bipolar central structure (low-density component) and spiral filaments (high-density component), the regions defining the inner and outer radii are not clear. We approximately measure $r_\mathrm{in}$ at the average of major and minor axis of the central structure and use the same $r_\mathrm{in}$ value for the filaments. We kept the $r_\mathrm{out}$ values of both the components as free parameters and varied within a range such that the model nebular size remained within the same order of the observed nebular extent. After computing a large number of models, we adopted a distance of 5.0 kpc, within the literature value. We estimated a luminosity of 6700 $L_{\sun}$. We initially adopt filling factors, $f_\mathrm{h}=1.0$ and $f_\mathrm{l}=0.5$ for the high- and low-density components in the model, respectively, and slightly varied close to those values. Finally, the values remained unaltered in the best-fitting model, hence, obtaining $f_\mathrm{h}=1.0$ and $f_\mathrm{l}=0.5$. 

Theoretically, these model estimated values are degenerate with another model with the parameters, $d=x\times5.0$ kpc, $L=x^2\times6700$ $L_{\sun}$ and $f_\mathrm{h}=1.0/x$ and $f_\mathrm{l}=0.5/x$. From \citet{1994ApJS...92..125V} we used the values computed for `H-burning PNN evolutionary models' ($Z=0.016$) and find a relation between the age ($t$) and the luminosity ($L$) of the progenitor for $T_\mathrm{eff}=92000$ K as,  
\begin{equation}
\mathrm{Log}(t)=12.4599-2.36256\times\mathrm{Log}(L/L_{\sun})
\end{equation}
We assume the nebular expansion velocity, $v_\mathrm{exp}=50$ km s$^{-1}$, obtained for the central structure in the 3D model, we obtain a nebular age of 1421 yr, considering $r_\mathrm{out}=0.073$ as the edge of the nebula. Thus, for the degenerate model, we have $t=x\times3700$ yr, along with $L/L_{\sun}=x^2\times6700$, leading to the relation,
\begin{equation}
\mathrm{Log}(t)=1.24+0.5\times\mathrm{Log}(L/L_{\sun})
\end{equation}
Solving equations 6 and 7, we obtain a luminosity of 8309 $L_{\sun}$, with $x=1.1136$, thereby giving, $d=5.57$ kpc, $f_\mathrm{h}=0.9$ and $f_\mathrm{l}=0.45$. For $d=5.57$ kpc, the value of inner radii for both the nebular components to compute the photoionization model are set at $r_\mathrm{in}=0.029$ pc.

\subsection{Progenitor Mass}

\begin{figure}
\centering
\scalebox{0.35}[0.35]{\includegraphics{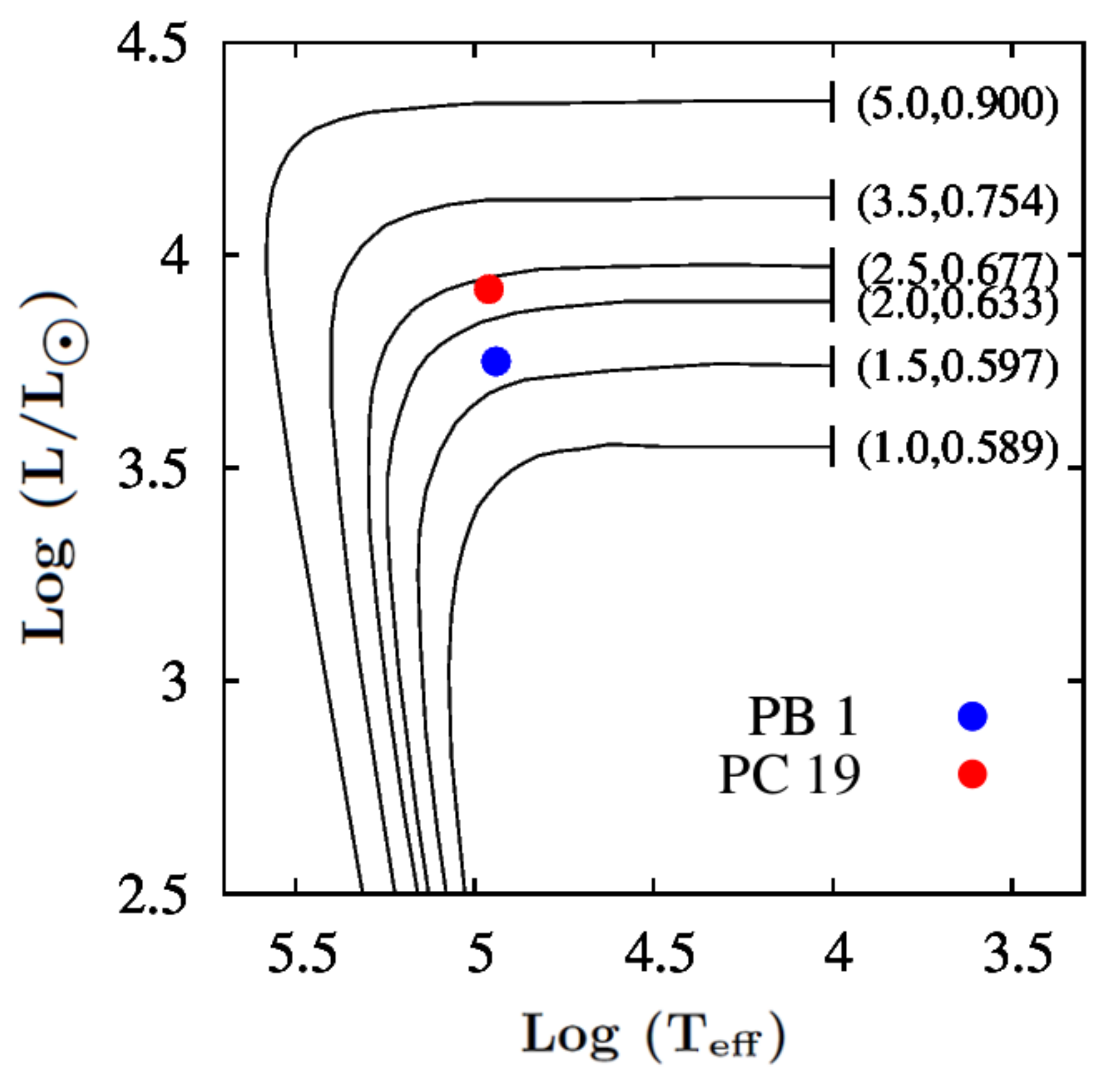}}
 \caption{Plot of Log $(L/L_{\sun})$ vs. Log $(T_\mathrm{eff})$. The line plots shown are the post-AGB model tracks taken from Vassiliadis $\&$ Wood (1994) ($Z=0.016$). The label at the beginning of each track corresponds to initial (main-sequence) mass, and core mass at Log $(T_\mathrm{eff})=4$, of the star in terms of solar units. Positions of PB 1 and PC 19 in the plot are marked with points.}
 \label{fig:massfunction}
\end{figure}

We estimate the mass of the central star of the PN, using H burning post-AGB model tracks (Log $(L/L_{\sun})$ vs. Log $(T_\mathrm{eff})$) (corresponding to $Z=0.016$) from \citet{1994ApJS...92..125V} (Figure \ref{fig:massfunction}). They considered mass-loss based on empirical mass-loss rates of PNe progenitors and radiation-pressure-driven stellar wind theory, and calculated models for different initial masses and metallicities. Each track is labelled in the format (X, Y), where, `X' corresponds to the value of initial (main-sequence) mass and `Y' to the core mass at Log $(T_\mathrm{eff})=4$, in terms of solar masses. We place the two PNe on the model tracks at the coordinates (Log $(L/L_{\sun})$, Log $(T_\mathrm{eff})$) corresponding to our estimated values of temperature and luminosity of the central stars from photoionization modelling. By interpolating, we estimate masses of $\sim$1.67 and $\sim$2.38 ${M_{\sun}}$ for the progenitors of PB 1 and PC 19, respectively.

\section{Summary and Discussion} \label{sec:discuss}

In this paper, we present the result of our study of two less-studied PNe, PB 1 and PC 19. For both the PNe, we study the morphology and ionization characteristics by computing models that are satisfactorily constrained by the observations. For each PN, we reconstruct 3D morphology from the 2D image using SHAPE and compute photoionization models by fitting the observed spectra using CLOUDY. We find that the projection of the 3D morphology matches well with the observed image. Our best-fitting photoionization models match the  observed spectra, including the continuum, absolute H$\beta$ flux and the relative line fluxes well. The r.m.s. values of the models of both PB 1 and PC 19 are well below unity, which reflects the goodness of fit of the models. The assumed shell geometry used in the photoionization models are guided by the morphological information derived from the 3D modelling. This improves the fitting of the photoionization models significantly. We compute a photoionization model for PC 19, without considering any shock excitation scenario or overabundance of N for the low ionization components. Several physical parameters are obtained self-consistently from the models. From the 3D models, we estimate the relative dimensions of the structural components and their orientation angles and expansion velocities. We estimate temperature, luminosity and gravity of the central star and nebular elemental abundances from the photoionization models. We estimate distance to the PNe as $\sim$4.3 kpc for PB 1 and $\sim$5.6 kpc for PC 19.   

According to our 3D model, PB 1 consists of a point symmetric inner shell, surrounded by a bipolar halo. Our morphological study reveals that the stellar wind might had directionality since the early phases of post-AGB evolution, as evident from the bipolarity of the halo. The overall nature of the inner shell structure obtained from the 3D analysis is physically similar with the apparent 2D morphology. Both of them imply a point symmetric outflow of the fast wind. The morphology of PC 19 consists of a central structure in an open lobe bipolar morphology with a pair of filamentary arms. The spiral arms are contained within the lobes of the nebula. The top view of the model of PC 19 seems to have similarity with image of NGC 6881 (SMV11) as observed from Earth. We obtain the velocity field in the filamentary arms of PC 19 from the PV diagrams. While the blueward arm shows a monotonically increasing velocity profile, the profile for the redward arm depicts a sudden rise in the outer region of the filaments. 

The aim of our 3D modelling approach in this work has been to obtain a 3D structure with possible components that would reproduce the observed image satisfactorily. We do not attempt to estimate the parameters, such as, dimensions, expansion velocity, position and inclination angles of the components, very accurately as that would had been subjected to the degeneracy while measuring the parameters simultaneously. The comparison of the observed and the modelled images are done with eye-estimation. However, the models might be subject to a collective error in the range of $\sim$15-20$\%$, estimated by noting deviations of the modelled images that closely match the observed image.     

Overall, our photoionization model results match with the observations. In our modelling approach, large number of models were generated by varying the parameters, the best-fitting model was chosen subsequently. However, degeneracy may occur for a different set of model parameters. This might be resolved using independent information about the parameters and comparison with more model spectra generated by varying the parameters in smaller steps. It is very complicated process to estimate the error, more precisely, associated with individual model parameters. By running large number of models, changing the parameters around their estimated values, and noting deviation of the observables from the acceptable range, the error associated with the parameters could be calculated.

Further, during observations, the slit was placed along the central region of the nebulae (as shown in Figures \ref{fig:3Dmodelpb1} and \ref{fig:3Dmodelpc19}). This might have led to some consequences on the results. As we consider a spherical photoionization model to reproduce the observations, the emission from outer region might have been overweighted compared to the observations. Moreover, within the nebula, higher ionization states are distributed towards the central region and lower ionization ones towards the outer regions. As a result, the He~{\sc ii} lines might have been enhanced in the observed spectrum and led to a higher estimation of $T_\mathrm{eff}$. Also, the discrepancies between the ICFs calculated from photoionization modelling and those calculated from direct method might be due to this aperture effect (\citealt{2017IAUS..323...43M}), which has led to discrepancies between model abundances and the abundances estimated using ICFs. Interestingly, for N, O and Ne abundances, we find that this discrepancy is lower in PC 19 than in PB 1. This is in line with the fact that higher percentage of nebula is covered by the slit during our observation of PC 19 compared to PB 1. More observations at different slit positions and a 3D photoionization model (e.g., \citealt{2016MNRAS.457.3409A}; \citealt{2016A&A...585A..69G}) may help to yield more precise results.

The higher abundances of elements, particularly of He, C and N, in PN indicate towards higher progenitor mass and bipolar morphology (e.g., \citealt{2002ApJ...576..285S}, \citealt{2003MNRAS.340..883P}). The abundance, C/H, estimated for PB 1 in this work, is quite high, He/H and N/H are also towards the higher side. This is in line with the result that the estimated progenitor mass for PB 1 $\sim$1.67 $M_{\sun}$ is towards the high-mass range (starting from $\sim$2 $M_{\sun}$) found in literature. In case of PC 19, the morphology is highly bipolar. We obtain a progenitor mass of $\sim$2.38 ${M_{\sun}}$, which is above the high-mass range of PN progenitors. However, the estimated abundances of He, C, N and O in PC 19 are lower than their average values found in bipolar PNe \citep{2003MNRAS.340..883P}. Further investigation is needed to resolve this contradiction.

As the reliability of a model depend upon the constrains available through observations, there are much scope of improvements through observational perspective. More PV diagram in various orientation of the slits are needed to constrain the morphology with confidence. Deep high-resolution spectra of the objects may help to estimate weak line fluxes and measure the electron temperatures and densities precisely. More detailed morphologies to represent the non-spherical structures such as spiral, bipolar lobes, etc., in place of simplified geometries could be used as the input for the photoionization models, which may improve the overall balance of different ionization states in the model.

\section*{Acknowledgements}   

We are thankful to the reviewer, Dr. Christophe Morisset, for his critical and valuable comments that helped to improve this paper. We acknowledge S. N. Bose National Centre for Basic Sciences under Department of Science and Technology (DST), Govt. of India, for providing necessary support to conduct research work. We are thankful to the HCT Time Allocation Committee (HTAC) for allocating nights for observation, and the supporting staff of the observatory. This paper uses data based on observations made with the NASA/ESA Hubble Space Telescope, and obtained from the Hubble Legacy Archive, which is a collaboration between the Space Telescope Science Institute (STScI/NASA), the Space Telescope European Coordinating Facility (ST-ECF/ESA) and the Canadian Astronomy Data Centre (CADC/NRC/CSA). We also use data from IAC Morphological Catalog of Northern Galactic Planetary Nebulae, and San Pedro M\'artir Kinematic Catalogue of Galactic Planetary Nebulae. We thank the people maintaining these data bases.

\bibliographystyle{mnras}
\bibliography{References}

\bsp	
\label{lastpage}
\end{document}